\documentclass[12pt,preprint]{aastex}

\slugcomment{Not to appear in Nonlearned J., 45.}

\shorttitle{AGN AND HOST GALAXY CO-EVOLUTION FROM HARD X-RAY EMISSION}
\shortauthors{Wang et al.}

\begin{document}


\title{INSIGHTS INTO AGN AND HOST GALAXY CO-EVOLUTION FROM HARD X-RAY EMISSION}


\author{J. Wang\altaffilmark{1} X. L. Zhou\altaffilmark{1} and J. Y. Wei\altaffilmark{1}}
\email{wj@bao.ac.cn}
\altaffiltext{1}{National Astronomical Observatories, Chinese Academy of Sciences}







\begin{abstract}
We study the AGN-host co-evolution issue here by focusing on the correlation between
the hard X-ray emission from central AGNs and the stellar populations of the host galaxies. 
By focusing on the galaxies with strong H$\alpha$ line emission (EW(H$\alpha)>5$\AA),
both X-ray and optical spectral analysis are
performed on 67 (partially) obscured AGNs that are selected from 
the \it XMM-Newton\rm\ 2XMMi/SDSS-DR7 catalog originally cross-matched by Pineau et al.
The sample allows us to study central AGN activity and host galaxy directly and simultaneously in individual objects. 
Combining the spectral analysis in both bands reveals that 
the older the stellar population of the host galaxy, the harder the X-ray emission will be,
which was missed in our previous study where the \it ROSAT\rm\ hardness ratios are used.
By excluding the contamination from the host galaxies and from the jet beaming emission, the 
correlation indicates that the Compton cooling in the accretion disk corona decreases with the 
mean age of the stellar population. We argue that the correlation is related to the correlation 
of $L/L_{\mathrm{Edd}}$ with the host stellar population. In addition,
the [\ion{O}{1}]/H$\alpha$ and [\ion{S}{2}]/H$\alpha$ narrow-line ratios are identified to correlate with 
the spectral slope in hard X-ray, which can be inferred from the currently proposed evolution of the X-ray emission
because of the confirmed tight correlations between the two line ratios and stellar population age.
 \end{abstract}


\keywords{galaxies: nuclei - galaxies: evolution - X-rays: galaxies}



\section{INTRODUCTION}

The co-evolution of active galactic nuclei (AGNs) and their host galaxies has been a hot 
topic in astrophysics for many years. The co-evolution issue is basically implied by several
observational facts. At first, the mass of a central supermassive blackhole (SMBH) is found to be strongly correlated 
with various properties of the host galaxy where the SMBH resides in. 
These properties include the velocity dispersion, luminosity and mass of the bulge
of the host galaxy (e.g., Magorrian et al. 1998; Tremaine et al. 2002;
Ferrarese \& Merritt 2000; Greene \& Ho 2006; Greene et al. 2008; Haring \& Rix 2004;
Gebhardt et al. 2000a, b; Merritt \& Ferrarese 2001; McLure \& Dunlop 2002; Woo \& Urry 2002; Woo et al. 2010;
Ferrarese \& Ford 2005; Gultekin et al. 2009).    
Secondly, the cosmic star formation history is found to be traced by the evolution of number density
of Quasi-stellar objects (QSOs), and by the growth of central SMBH since $z\sim5$ to the present
(e.g., Nandra et al. 2005; Silverman et al. 2008; Shankar et al. 2009; Hasinger et al. 2005; Ueda et al. 2003; Croom et al.
2004; Aird et al. 2010; Assef et al. 2011). Thirdly, about half of local AGNs are identified to be associated 
with circumnuclear young stellar populations and violent ongoing star formation activities (see Wang \& Wei 2010 for a brief summary of the 
references). Finally, both AGNs and starforming galaxies show the ``down-sizing'' effect 
in which both activities at later epoch predominantly occur in the less massive systems with lower luminosities 
 (e.g., Cowie et al. 2003; Bongiorno et al. 2007; Ueda et al. 2003; McLure \& Dunlop 2004; Hasinger et al. 2005; 
Alonso-Herrero et al. 2008).

With these implications, significant observational progress has been achieved in the past decade to understand the 
elusive evolutionary relationship between the star formation and SMBH growth. Heckman \& Kauffmann (2006 and 
references therein) indicates that the majority of local SMBH growth occurs not only in high accretion phase, 
but also in the galaxies with young stellar populations and with ongoing or recent star formations. This conclusion is
recently extended to less massive SMBH by Goulding et al. (2010). 
Not only the theoretical simulations, but also the analysis of AGN's host galaxies
suggest a possible delay of $\sim10^2$Myr for the detectable AGN  phenomenon after the onset
of star formation activity (e.g., Wang \& Wei 2006; Schawinski et al. 2007; 
Reichard et al. 2008; Li et al. 2008; Wild et al. 2010; Zhou et al. 2005, Davis et al. 2007; Hopkins et al. 2005). 
Moreover,  as compared to the distribution of quiescent galaxies, 
observations show that there is an excess of AGNs 
in the ``green valley'' between the blue cloud and red sequence, 
which is commonly interpreted by the suppression of star formation due to the feed back from AGNs 
(e.g., Sanchez et al. 2004; Nandra et al. 2007; Schawinski et al. 2009; Treister et al. 2009).


A natural question on the complicated co-evolution issue is how AGNs co-evolve with their host galaxies, and
which parameter of AGN's accretion determines the coevolution. In order to answer the question, one needs to 
study both accretion activity and host galaxy properties simultaneously in individual AGN.   
It is well known that AGNs are luminous X-ray emitters up to 100keV. The hard X-ray emission is 
a powerful tool to identify nuclear SMBH accretion activity and to study the accretion process that fuels AGNs,
both because it can penetrate the obscuration material much more easily than lower energy emission and
because it is produced in the region very close to SMBH.
Wang et al. (2010) studied a sample of \it ROSAT\rm-selected partially obscured AGNs to explore 
the evolutionary role of the X-ray emission. 
Although the X-ray spectral slopes are found to be roughly correlated with
the two narrow emission-line ratios, i.e., [\ion{O}{1}]/H$\alpha$ and [\ion{S}{2}]/H$\alpha$,
the authors failed in identifying a correlation between  
the slopes and host galaxy stellar population ages. They argued that the nondetection of the correlation is mainly due to
the large uncertainties of the used X-ray spectral slopes.
The slopes are 
estimated from the \it ROSAT\rm\ hardness ratios, and the intrinsic absorptions are entirely ignored in their study.
To overcome this fundamental drawback, we perform a systematic spectral study on a sample selected from 
the \it XMM-Newton\rm\ 2XMMi catalog cross-correlated
with Data Release 7 of the Sloan Digital Sky Survey (SDSS) to explore the 
co-evolution of the AGN's X-ray emission and AGN's host galaxy. The cross-correlation was originally done by 
Pineau et al. (2011). 
In the current study, we mainly focus on the type II and partially obscured AGNs because the starlight 
component from host galaxy and the narrow emission-lines can be easily and accurately extracted from individual observed optical spectrum.     
Meanwhile, it is emphasized that the X-ray spectral slope can be highly improved for individual 
object through the spectral fitting on the wider energy band. 
The \it XMM-Newton\rm\ satellite (Jansen et al. 2001) was launched by the European Space Agency in
1999. The instruments onboard the satellite works in the energy range from 0.2 to 12 keV.

The paper is organized as follows.
\S2 presents the sample selection and data reductions, including the starlight component removal,
line profile modeling and X-ray spectral fitting. The analysis and results are shown in
\S 3. The discussion is presented in \S 4.   
A $\Lambda$ cold dark matter ($\Lambda$CDM) cosmology with parameters $h_0$ = 0.7, $\Omega_0$ = 0.3, 
and $\Omega_\Lambda$ = 0.7 (Spergel et al. 2003) is adopted throughout the paper.

\section{SAMPLE SELECTION AND DATA REDUCTION}
\subsection{Sample Selection From the 2XMMi/SDSS-DR7 Catalog}

The incremental Second \it XMM-Newton\rm\ Serendipitous Source Catalog (2XMMi) is an updated version of the 
2XMM catalog (Watson et al. 2009). The 2XMMi catalog contains a total of 221,012 unique, serendipitous X-ray sources, and is the 
largest catalog of X-ray sources ever published at that time. Because of the high throughput of the 
EPIC cameras onboard the \it XMM-Newton\rm\ satellite, the catalog reaches a 90\% completeness at 
a sensitive of $1\times10^{-14}\ \mathrm{erg\ s^{-1}\ cm^{-2}}$ and $9\times10^{-14}\ \mathrm{erg\ s^{-1}\ cm^{-2}}$
in the 0.5-2.0 keV and 2.0-12.0 keV bandpass, respectively. An accuracy of 2\arcsec\ is typical for the source positions.   
Pineau et al. (2011 and references therein) recently performed a 2XMMi/SDSS program to identify the optical counterparts
of the \it XMM-Newton\rm\  X-ray sources from the SDSS-DR7 catalog (Abazajian et al. 2009). The cross-correlation uses the  
traditionally adopted likelihood ratio estimator that is only based on the probability of spatial coincidence of the 
X-ray source and the optical candidate. By only considering the point-like sources with a position accuracy $\leq5\arcsec$ in X-ray, 
the cross-correlation returns a total of more than 30,000 X-ray sources that have a SDSS-DR7 optical counterpart  with
an identification probability larger than 90\% (Pineau et al. 2011).

A sub-sample is selected from the 2XMMi/SDSS-DR7 catalog as follows.
At first, to exclude spurious matches as far as possible, we require that a) the probability of identification is no less than 95\%; 
b) the angular distance between individual \it XMM\rm -newton X-ray source and the corresponding optical counterpart is not larger than 3\arcsec, 
taking into account of the SDSS fiber aperture\footnote{The middle panel in Figure 3 in Pineau et al. (2011) shows that 
there are a few of matched objects whose matching distances are larger than 5\arcsec, even when the identification probability is larger than 
95\%. The aperture selection criterion is useful to ensure that the detected X-ray emission mainly comes from the center of each 
galaxy because the SDSS fiber is located on the center of each galaxy by the SDSS photometric analysis.};
c) the redshift is smaller than 0.2; d) the objects are not classified as star, 
according to the spectral type classification given by the SDSS pipelines (Glazebrook et al. 1998; Bromley et al. 1998).
There are 1327 objects fulfilling the above criteria. 
In order to avoid the effect caused by the optical spectra with poor quality, 
we further require that the $g$-band brightness is brighter than 19 mag. 
An equivalent width of the H$\alpha$ emission line larger than 5\AA\ is necessary to exclude BL\,Lac objects and quiescent galaxies (e.g.,
the so-called X-ray Bright Optically Normal Galaxies, XBONGs) as 
far as possible (e.g., Caccianiga et al. 2008), because one can not ensure that the X-ray emission from the quiescent galaxies (especially
the elliptical galaxies) do come from the central AGNs in the latter case (e.g., Fabbiano 1989).  With the two selection rules,
we are left with 733 sources. 
The objects with X-ray flux in the 0.2-12 keV bandpass lower than $1\times10^{-14}\ \mathrm{erg\ s^{-1}\ cm^{-2}}$
are removed from the subsequent spectral modelings, which 
excludes the extracted X-ray spectra with low  photon count rates.


There is a total of 362 2XMMi/SDSS-DR7 objects fulfilling the above selection criteria, after removing some duplications. 
Nine objects are at first removed from the subsequent spectral modeling because of the poor sky-line 
subtraction at the H$\beta$ region. In addition, the SDSS optical spectra show that: SDSS\,J122153.96+042742.5 is a \ion{H}{2} region 
within NGC\,4303;
SDSS\,J121901.36+471524.9 (NGC\,4258) is a typical starforming galaxy associated with a Wolf-Rayet bump at 
\ion{He}{2}$\lambda4686$; SDSS\,J093249.57+472522.8 shows a spectrum typical of a CV star, and is
incorrectly classified as a QSO by the SDSS pipelines.  
Among the remained 350 sources, there are 268 narrow emission-line galaxies (almost all the 
objects are classified as either transition galaxies or type II AGNs, see Section 3.2 below) and 82 type I AGNs, according
to the classification done by the SDSS pipelines.  
The SDSS spectra of the type I AGNs are then inspected one by one by eyes.
The inspection yields that 19 out of the 82 objects show a continuum that is dominated by the starlight from their host galaxies
(hereafter partially obscured AGNs for short).
The other 14 objects can be classified as Seyfert 1.5 galaxies in which the spectra are dominated by the featureless
continuum from the AGNs and show evident narrow emission lines from AGNs. The rest are broad-line AGNs and narrow-line 
Seyfert 1 galaxies. 

With these selection procedures,
the sub-sample used in the subsequent spectral analysis therefore contains the 268 narrow emission-line galaxies,
the 19 partially obscured AGNs and the 14 Seyfert 1.5 galaxies\footnote{The Seyfert 1.5 galaxies are useful in 
studying the relationship between the narrow emission-line ratios and their hard X-ray emission (see \S3.3.2 for the details),
because in these galaxies the narrow component of the Balmer lines can be accurately extracted from the observed line profile.}.

\subsection{SDSS Optical Spectroscopy}

The 1-Dimensional optical spectra listed in the sub-sample are analyzed by the IRAF\footnote{IRAF is distributed by  National Optical Astronomy Observatory, which
is operated by the Association of Universities for Research in Astronomy, Inc.,
under cooperative agreement with the National Science Foundation.} package through the standard procedures as follows.
For each of the spectra, the Galactic extinction is at first corrected by the color excess, the parameter
$E(B-V)$ taken from the Schlegel, Finkbeiner, and Davis Galactic
reddening map (Schlegel et al. 1998), by assuming an $R_V=3.1$ extinction law of milky way
(Cardelli et al. 1989). The spectrum is then transformed to the
rest frame, along with the flux correction due to the relativity effect, given the
redshift provided by the SDSS pipelines. 

For the narrow emission-line galaxies and partially obscured AGNs, 
the stellar absorption features
are subsequently separated from each rest-frame spectrum by
modeling the continuum and absorption features by the sum of
the first seven eigenspectra (see \S4.4 for the discussion of the contamination from the 
scattered AGN's continuum). The eigenspectra are built through the 
principal component analysis (PCA) method (e.g., Li et al. 2005; Hao et al. 2005; 
Wang \& Wei 2008; Boroson \& Lauer 2012; Francis et al. 1992) from the
standard single stellar population spectral library developed by
Bruzual \& Charlot (2003).
A Galactic extinction curve with $R_V=3.1$ is adopted in the 
modeling to account for the intrinsic extinction due to the host galaxy.   
A powerlaw continuum and an empirical \ion{Fe}{2} template
are additionally required to appropriately fit the underlying continuum in five objects. 
The powerlaw index is fixed to be the typical value of type I AGNs 
(i.e, $f_\lambda\propto\lambda^{-1.7}$, Vanden Berk et al. 2001 and references therein) in the fitting. 
The empirical \ion{Fe}{2} template given by Boroson \& Green
(1992) is used in the continuum modeling, after the template is broadened to the FWHM of the 
H$\beta$ broad component by convolving with a Gaussian profile.

For the Seyfert 1.5 galaxies, we model the underlying continuum by a combination of a broken powerlaw and 
the \ion{Fe}{2} template, because the continuum is dominated by the emission from central AGNs.

A $\chi^2$ minimization is performed for each of the spectra over
the rest-frame wavelength range from 3700 to 8000\AA, except for the regions with strong emission lines
\footnote{The wavelength regions used in the spectral modelings are $\lambda\lambda$3700-3717, 
$\lambda\lambda$3740-3860,$\lambda\lambda$3880-4335,
$\lambda\lambda$4345-4675,$\lambda\lambda$4690-4850,$\lambda\lambda$4870-4950, $\lambda\lambda$5020-6290,$\lambda\lambda$6310-6540,$\lambda\lambda$6600-6710,$\lambda\lambda$6740-8000.}, 
they are 
H$\alpha$, H$\beta$, H$\gamma$, [\ion{O}{3}]$\lambda\lambda$4959, 5007, \ion{He}{2}$\lambda$4686, [\ion{N}{2}]$\lambda\lambda$6548, 6583,
[\ion{S}{2}]$\lambda\lambda$6716, 6731, [\ion{O}{3}]$\lambda$4363, [\ion{O}{2}]$\lambda$3727, [\ion{Ne}{3}]$\lambda\lambda$3869, 3967, 
and [\ion{O}{1}]$\lambda$6300.

After removing the continuum from each observed spectrum, the emission-line profiles 
are modeled by the SPECFIT task (Kriss 1994) in the IRAF package for both H$\alpha$ and H$\beta$
regions, they are in the wavelength ranges: $\lambda\lambda$6500-6750 and $\lambda\lambda$4820-5050. 
Each line profile is modeled by a linear combination of a set of several Gaussian profiles. 
A broad H$\alpha$ or H$\beta$ is required in the modelings if the observed profile can not be properly fitted by the narrow component alone. 
The intensity ratios of the [\ion{O}{3}] and [\ion{N}{2}] doublets are fixed to their theoretical values.
The flux of the [\ion{O}{1}]$\lambda$6300 emission line is measured through direct integration by the SPLOT task in 
the IRAF package.

\subsection{XMM-Newton EPIC Spectra}  

Only the \it XMM-Newton\rm\ EPIC PN (Struder et al. 2001)
data are used in our X-ray spectral analysis. The data are reduced by the SAS v11.0 
software\footnote{http://xmm,esac.esa.int/} and by the corresponding calibration files. 
For each of the objects listed in the sub-sample,
the events corresponding to patterns 0-4 are selected from the PN data, and the CCD chip gaps are avoided. 
The bad and hot pixels are then removed from the original image.
The source spectrum is extracted from a circular aperture at the detected source position.
The aperture has a radius of 25-40\arcsec\, 
depending on the brightness of the object. The background is obtained from a circular source-free region that 
is offset from but close to the source. The 
pile-up in the data is checked by the SAS task \it epatplot\rm. 
The tasks \it rmfgen\rm\ and \it arfgen\rm\ are used to generate the needed response files. 

There are 92 objects that have adequate photon count rates for the next spectral modelings. 
The extracted spectra are fitted by the XSPEC package (Arnaud 1996). The fittings are performed via a basic 
model\footnote{A powerlaw photon spectrum is defined as $N(E)\propto E^{-\Gamma}$, where $E$ is the photon energy and 
$\Gamma$ is the photon index.} expressed as
$wabs*zwabs*powerlaw$ over the 0.2-8 keV band for most of the objects, because of their relatively low 
photon count rates. The neutral reflection model ($pexrav$, Magdziarz \& Zdziarski 1995) is required in four objects
to reproduce the ``bump'' at the high energy end (e.g., Zycki et al. 1994). We add an additional Gaussian profile
in eight sources to reproduce the narrow iron K$\alpha$ emission lines at 6.4 keV (rest frame).
SDSS\,J030349.10-010613.4 (NGC\,1194) shows the most complex X-ray energy spectrum among the sub-sample. 
In the object, we add a $mekal$ component with $kT\sim1$keV to match better the emission below 2keV. 
In addition to the Fe K$\alpha$ emission line, the spectrum shows the K$\alpha$ emission lines from Si, S, and Ca that are
superposed on a neutral reflection component (e.g., Greenhill \& Tilak 2008). The spectral fitting of each object includes an absorption due to 
the Galactic column density at the line-of-sight. The value of density is
taken from the Leiden/Argentine/Bonn (LAB) Survey (Kalberla et al. 2005). As an illustration, the best-fitting model
is reproduced in Figure 1 for SDSS\,J151640.21+001501.8.





\section{RESULTS AND ANALYSIS}

In order to ensure the reliability of the inferred conclusions, only the objects whose $1\sigma$ uncertainties 
of the X-ray photon index are less than 0.3 are 
considered in the subsequent statistical analysis. Finally, there are 67 objects
fulfilling the criterion (hereafter final sample for short). 
The results derived from the above spectral analysis are tabulated in Table 1 for the final sample.
The identification of each object and the corresponding redshift are listed in Columns (1) and
(2), respectively. Columns (3) to (6) tabulate the line ratios of
[\ion{N}{2}]$\lambda6583$/H$\alpha$, [\ion{S}{2}]$\lambda\lambda6726,6731$/H$\alpha$,
[\ion{O}{1}]$\lambda6300$/H$\alpha$, and [\ion{O}{3}]$\lambda5007$/H$\beta$ in logarithm
for each of the objects. The fitted spectral photon index $\Gamma_\mathrm{2-10keV}$ in hard X-ray is listed 
in Column (8). All the quoted errors correspond to a 1$\sigma$ significance level.   
Columns (9) and (10) show the spectral fitting inferred intrinsic X-ray luminosity 
in the 2-10keV bandpass $L_{\mathrm{2-10keV}}$ and the intrinsic [\ion{O}{3}] line luminosity ($L_{\mathrm{[OIII]}}$),
respectively. The line luminosity
is corrected for the local extinction. The extinction is inferred from the narrow-line ratio 
H$\alpha$/H$\beta$ by assuming
the Balmer decrement for standard case B recombination
and the Galactic extinction curve with $R_V=3.1$.

\subsection{Stellar Population Age Measurements}
The 4000\AA\ break index $D_\mathrm{n}(4000)$ (Bruzual 1983; Balogh et al. 1999) defined as
\begin{equation}
D_\mathrm{n}(4000) = \frac{\int^{4100}_{4000}f_\lambda d\lambda}{\int^{3950}_{3850}f_\lambda d\lambda}
\end{equation}
is widely used as an age indicator of the stellar population of the bulge of a 
galaxy (e.g., Heckman et al. 2004; Kauffmann et al. 2003c;
Kauffmann \& Heckman 2009; Kewley et al. 2006; Wang \& Wei
2008, 2010; Wild et al. 2007, 2010). It is believed that $D_\mathrm{n}(4000)$ is an excellent 
mean age indicator until a few Gyr after the onset of a star formation activity (e.g.,
Kauffmann et al. 2003c; Bruzual \& Charlot 2003). 
We measured the $D_\mathrm{n}(4000)$ index in the 
removed starlight spectra for all of the 268 narrow emission-line galaxies and
the 19 partially obscured AGNs. The measured values are tabulated in Column (7) in Table 1 for the objects listed in
the final sample. By using the same 
data reduction method described above, Wang et al. (2011) has estimated the typical 
uncertainty of $\sim0.03$ for the measured $D_n(4000)$ from the duplicate SDSS observations.

\subsection{BPT Diagnostic Diagram}
Figure 2 displays the [\ion{N}{2}]$\lambda6583$/H$\alpha$ versus [\ion{O}{3}]$\lambda5007$/H$\beta$ Baldwin-Phillips-Terlevich (BPT) diagnostic
diagram for the final sample. The BPT diagram was originally proposed by 
Baldwin et al. (1981), and then refined by Veilleux \& Osterbrock (1987). The diagram is commonly 
used to determine the dominant powering source in narrow emission-line galaxies
through their emission-line ratios. AGNs are concentrated in the upper right corner of the BPT diagram
because of their harder ionizing fields.
We plot the narrow emission-line galaxies and the 
broad-line AGNs (i.e., the partially obscured AGNs and Seyfert 1.5 galaxies) in Figure 2 by the filled and open squares, respectively.
The broad-line emission that is obtained through our profile modeling is excluded from the line ratio calculations.
The solid line shows the empirical demarcation line that is 
proposed by Kauffmann et al. (2003) to separate `pure' starforming galaxies according to the large sample
provided by SDSS. Stasinska et al. (2006) proposed a theoretical demarcation line between 
`pure' starforming galaxies and AGNs based on their photoionization models. The theoretical
line is shown by the dashed line in Figure 2, which is close to and slightly restrictive than the empirical line drawn by
Kauffmann et al. (2003). Stasinska et al. (2006) argued that the Kauffmann line includes the starforming galaxies 
that have an AGN contribution to H$\beta$ up to 3\%.
As shown by the diagram, all the objects listed in the final sample, except one, are 
located above both demarcation lines.

\subsection{Statistical Analysis}
\subsubsection{X-ray Luminosity and Spectral Photon Index}

The left and right panels in Figure 3 show the distributions of $\Gamma_{\mathrm{2-10keV}}$ and $L_{\mathrm{2-10keV}}$ for the final sample,
respectively. The measured $\Gamma_{\mathrm{2-10keV}}$ has an average (median) value of 1.87 (1.83).
The values are highly consistent with the previous studies that indicate a typical index of  $\Gamma\sim1.9$
for radio-quiet AGNs (e.g., Zdziarski et al. 1995; Reeves \& Turner 2000; Piconcelli et al. 2005; Dadina 2008; Panessa et al. 2008; 
Zhou \& Zhang 2010; Corral et al. 2011; Panessa et al. 2008; Mateos et al. 2010).    
The 2-10keV luminosity spans from $10^{41}$ to $10^{44}\ \mathrm{erg\ s^{-1}}$, which is typical for 
type II AGNs (e.g., Heckman et al. 2005; Panessa et al. 2006; Singh et al. 2011). The average and median values of $L_{\mathrm{2-10keV}}$ are $10^{42.6}$ and 
 $10^{42.7}\ \mathrm{ergs\ s^{-1}}$, respectively. It is noted that the X-ray luminosities of
most known X-ray luminous star-forming and elliptical galaxies
are not higher than $L_X = 10^{42}\ \mathrm{ergs\ s^{-1}}$ (e.g., Zezas et al. 2003; Lira
et al. 2002a, 2002b; O'Sullivan et al. 2001).

\subsubsection{Correlations}

Figure 4 plots the tight correlation between $L_{\mathrm{2-10keV}}$ and $L_{\mathrm{[OIII]}}$. 
An unweighted fitting yields a relationship $\log L_{\mathrm{2-10keV}}=(0.97\pm0.11)\log L_{\mathrm{[OIII]}}+(2.69\pm4.53)$ 
with a 1$\sigma$ standard deviation of 0.63.
The tight correlation between $L_{\mathrm{2-10keV}}$ and $L_{\mathrm{[OIII]}}$ plays an important role in
estimating bolometric luminosities in type II AGNs (e.g., Heckman et al. 2004; Brinchmann
et al. 2004; Kewley et al. 2006). The correlation was reported in previous studies by many authors
(e.g., Heckman et al. 2005; Jin et al. 2012; LaMassa et al. 2011; Georgantopoulos \& Akylas 2010), although 
Heckman et al. (2005) indicates that the correlation becomes much weaker for an optically selected Type II AGN 
sample. They argued that the weaker correlation is mainly due to the intrinsic photoelectric absorption in X-ray band.
Recent studies, however, show an agreement of luminosity functions between the optically and X-ray-selected AGNs
(e.g., Georgantopoulos \& Akylas 2010; Trouille \& Barger 2010). The insert panel in Figure 4 presents the distribution
of the $L_{\mathrm{2-10keV}}/L_{\mathrm{[OIII]}}$ ratio. The distribution has a
mean value of 1.4, and a dispersion of 0.63 dex. Again, one can 
see a consistency with the previous studies (see the citations quoted above).

The main results of this paper are shown in Figure 5 for the final sample, in which the fitted $\Gamma_{\mathrm{2-10 keV}}$
is plotted as a function of $D_n(4000)$, [\ion{S}{2}]/H$\alpha$ and [\ion{O}{1}]/H$\alpha$ in the left, middle and right panels,
respectively. Although the size of the current sample is smaller than that used in Wang \& Wei (2010),
much improved correlations (especially 
the $\Gamma_{\mathrm{2-10keV}}$ versus $D_n(4000)$ correlation) can be clearly identified from the figure  
thanks to the better determination of  photon index.
The results of Spearman rank-order tests are summarized in Table 2. The calculated correlation coefficients and 
the corresponding probabilities of null correlation are listed in the first and second rows, respectively.

The new anti-correlation between $\Gamma_{\mathrm{2-10keV}}$ and $D_n(4000)$ shows that 
the X-ray spectra of AGNs are a function of the age of stellar population of the host 
galaxies: older the stellar population, 
harder the central X-ray emission will be.  
The current study additionally reproduces the anti-correlations between the X-ray photon index and the 
two line ratios. Two similar correlations were identified in Wang \& Wei (2010) by using the \it ROSAT\rm\ hardness ratios.
Wang \& Wei (2010) argued that the correlations are 
consistent with the previous theoretical calculations and observational results. High energy observations show that
dozens of LINERs with high [\ion{O}{1}]/H$\alpha$ ratios have hard, flat X-ray spectra (e.g., Flohic et al. 2006; Gliozzi et al. 2008; 
Rinn et al. 2005). In the theoretical ground,  
photoionization calculations indicate that a hard ionizing field with a powerlaw index $\alpha<1.4$ can 
produce the strong [\ion{O}{1}] line emission (i.e., $\log(\mathrm{[OI]/H\alpha})>-0.6$, Kewley et al. 2006).

Figure 5 naturally implies that the two narrow emission-line ratios are related with the stellar population ages. In fact, 
Wang \& Wei (2008, 2010) identified two tight correlations between $D_n(4000)$ and the two line ratios 
in both optically- and \it ROSAT\rm\ X-ray-selected AGN samples. 
The two correlations are re-examined and confirmed here in the \it XMM-Newton\rm\ selected sample. The 
two line ratios of [\ion{S}{2}]/H$\alpha$ and [\ion{O}{1}]/H$\alpha$ are plotted against 
$D_n(4000)$ in Figure 6 for the sub-sample. The filled circles mark  
the objects that are listed in the final sample.
Strong correlations can be clearly identified for both samples from the figure. 
Spearman rank-order tests are performed to show the significance of the correlations. The estimated correlation
coefficients are listed in Table 3. The value shown in the bracket for each entry is the probability
of null correlation. By combining the relationship with $\Gamma_{\mathrm{2-10keV}}$, we argue that 
the two correlations are possibly driven by the change of AGN's X-ray spectrum as a function of age of its host galaxy,
although the ``line-mixing'' effect caused by star formation could not be entirely excluded.

\section{DISCUSSION}

The role of X-ray emission in the AGN-host coevolution issue is explored by our spectral analysis in both
optical and X-ray bands of a sample selected from the 2XMMi/SDSS-DR7 catalog. With the much better 
determination of the X-ray spectral photon index, the analysis allows us to identify a strong 
anti-correlation between the photon index and stellar population age, which was missed in our previous study.
The correlation indicates that the X-ray emission from the central SMBH accretion activity plays an important role in 
the AGN-host coevolution issue, in which the hardness of X-ray spectrum increases with the host galaxy
stellar population age.

\subsection{X-ray Emission in AGNs}
Because the hard X-ray emission is believed to be produced in the inner region of an AGN, the photon index 
$\Gamma_{\mathrm{2-10keV}}$ in hard X-ray is therefore important in studying the accretion process and emission mechanism
in AGNs. The hard X-ray spectra of radio-quiet Seyfert galaxies are dominated by a powerlaw component with a mean 
photon index of $\Gamma\sim1.9-2.0$ (Mushotzky et al. 1980; Nandra et al. 1990; Page et al. 2005; Shemmer et al. 2005;
Vignali et al. 2005; Just et al. 2007), which is traditionally explained by the 
comptonization of the seed UV/soft-X-ray photons from the accretion disk by the energetic electrons in hot plasma cloud 
(e.g., Haardt \& Maraschi 1991; Zdziarski et al. 2000; Kawaguchi et al. 2001). Previous studies 
frequently indicate that the properties of X-ray emission is strongly related with the ones derived from optical spectroscopic observations. 
There is a strong anti-correlation between $\Gamma$ and 
FWHM of AGN's broad H$\beta$ emission line (e.g., Brandt et al. 1997; Leighly 1999; Reeves \&
Turner 2000; Shemmer et al. 2006, 2008; Zhou \& Zhang 2010; Jin et al. 2012).
Direct correlation analysis and PCA analysis
indicate that the fundamental  Eigenvector-I (EI) space is strongly correlated with the soft X-ray spectral index 
(e.g., Wang et al. 1996; Laor et al. 1997; Vaughan et al. 1999; Grupe 2004; Xu et al. 2003; Boroson \& Green 1992 
and see Sulentic et al. 2000 for a review). Briefly, the larger the photon index, the stronger the
\ion{Fe}{2} blends and [\ion{O}{3}] line emission will be.

It is now commonly believed that the EI space is physically driven by $L/L_\mathrm{Edd}$, 
where $L_{\mathrm{Edd}}=1.26\times10^{38}(M_{\mathrm{BH}}/M_\odot)\ \mathrm{ergs\ s^{-1}}$ (e.g., Boroson 2002).
Thanks to the great progress made in the reverberation mapping technique, empirical relationships have been well 
established in past decades to estimate the viral blackhole mass in a type I AGN from 
single epoch optical spectroscopic observation (see a recent review in Marziani \& Sulentic 2012 and references therein).
The calibrations
allow various authors to identify a correlation between $\Gamma$ and $L/L_{\mathrm{Edd}}$ in radio-quiet type I AGNs
(e.g., Grupe 2004; Desroches et al. 2009; Gierlinski \& Done 2004; Lu \& Yu 1999; Porquet et al. 2004; Wang et al. 2004; Bian 2005, 
Shemmer et al. 2006, 2008; Risaliti et al. 2009; Jin et al. 2012; Zhou \& Zhao 2010), although non-monotonic trends are suggested 
by Kelly et al. (2008) and Gu \& Cao (2009) when low-luminosity AGNs are included. 
The correlation implies that the accretion rate is related to the 
physical conditions in the hot corona producing the hard X-ray emission. 
Pounds et al. (1995) points out that a high $L/L_{\mathrm{Edd}}$
state tends to produce a steep soft X-ray spectrum.
A commonly accepted explanation for the correlation is that: the corona cools more efficiently due to the inverse Compton scattering when the 
disk flux irradiating the corona increases. The enhanced cooling then results in a 
soft, steep X-ray spectrum at high $L/L_{\mathrm{Edd}}$ state. 
Cao (2009) recently suggests that the
correlation could be explained by the disk-corona model 
in which the corona is heated by magnetic field reconnections.

\subsection{Co-evolution of AGNs and Their Host Galaxies}

The main results obtained in this paper indicate that the physical process occurring 
in the central SMBH accretion activity correlates with the host galaxy stellar population:
the strength of Compton cooling in the corona around the SMBH+disk system decreases with the mean age of stellar population of the host galaxy. 
Taking into account of the strong correlation between $\Gamma$ and
$L/L_{\mathrm{Edd}}$, the $\Gamma_{\mathrm{2-10kev}}-D_n(4000)$ relationship is likely to be related to 
the correlation between $L/L_{\mathrm{Edd}}$ and the host galaxy.

Wang et al. (2006) extended the EI space into middle-far-infrared colors $\alpha(60,25)$ by 
performing a PCA analysis on a sample of \it IRAS\rm-selected Seyfert 1.5 galaxies.
Because the color $\alpha(60,25)$ addresses the relative
importance of AGN activity and starburst activity, the extension therefore suggests an evolutionary role of 
the EI space, which naturally implies an evolution of the photon index with the host galaxy stellar population. 
The evolutionary role of $L/L_{\mathrm{Edd}}$ has been frequently suggested in observational ground in past a few years.
A direct correlation between $L/L_{\mathrm{Edd}}$ and stellar population age of host galaxy is established
by different approaches: using $L(\mathrm{[OIII]})/\sigma_*^4$ as a proxy of $L/L_{\mathrm{Edd}}$ in type II AGNs 
(e.g., Kewley et al. 2006; Wild et al. 2007\footnote{In subsequent work, Wild et al. (2010) show that the 
accretion rate versus stellar population age
correlation does not hold follwing a starburst.}; Kauffmann et al. 2007) and estimating $L/L_{\mathrm{Edd}}$ directly from 
the Balmer broad lines in partially 
obscured AGNs (Wang et al. 2008, 2010). In addition to the stellar population age, $L/L_{\mathrm{Edd}}$ is found 
to be related with the ongoing SFR, when the SFR is assessed by the near-infrared polycyclic aromatic
hydrocarbon (PAH) emission (e.g., Watabe et al. 2008; Woo et al. 2012; Imanishi \& Wada 2004). 
Chen et al. (2009) identified a tight correlation between 
$L(\mathrm{[OIII]})/\sigma_*^4$ and the specific SFR in Seyfert 2 galaxies. The authors
argue that supernova explosions might play a key role in the fueling of gas to central SMBH (see also
in the numerical simulation studies, e.g. Wada et al. 2009; Schartmann et al. 2010).

As a subclass of AGNs, narrow-line Seyfert 1 galaxies (NLS1s)\footnote{NLS1s are 
commonly defined as the AGNs with narrow Balmer broad emission lines, weak [\ion{O}{3}] lines and strong 
\ion{Fe}{2} complex, i.e., $\mathrm{FWHM_{H\beta}<2000\ km\ s^{-1}}$, [\ion{O}{3}]/H$\beta_{\mathrm{total}}<3$, and
\ion{Fe}{2}/$\mathrm{H\beta_{total}} > 0.5$, see Komossa (2008) for a recent review.}
are statistically clustered at one extreme end of the AGN's correlation space 
(e.g., Boroson \& Green 1992; Zamfir et al. 2008; Grupe 2004). These galaxies are believed to have 
less massive SMBH and higher $L/L_{\mathrm{Edd}}$ than do broad-line Seyfert 1 galaxies
(BLS1s, e.g., Boroson 2002; Collin \& Kawaguchi 2004). Steeper X-ray spectra are on average found in 
NLS1s than in BLS1s (e.g., Laor et al. 1994; Boller et al. 1996; Wang et al. 1996; Zhou \& Zhang 2010;
Puchnarewicz et al. 1992; Grupe 1996, 2004; Boller et al. 1996; Brandt et al. 1997;
Leighly 1999; Comastri 2000; Vaughan et al. 2001; Zhou et al. 2006), although relatively 
flat X-ray energy spectra are identified in a very small fraction of NLS1s (e.g., Zhou et al. 2006; 
Williams et al. 2004). NLS1s are expected to be ``young'' AGNs at early evolutionary
stage associated with intense circumnuclear star formation and with young stellar populations 
(see also in Mathur 2000 for other arguments). In fact,  Sani et al. (2010) recently revealed an enhanced star 
formation activity in NLS1s from the \it Spitzer\rm\ spectroscopy as compared with BLS1s with the same AGN luminosity.
The morphological study
indicates a higher fraction of nuclear star-formation rings for NLS1s than for BLS1s (Deo et
al. 2006). Based on the SDSS survey, Mao et al. (2009) found three particular broad-line AGNs whose narrow emission lines are 
identified to be ionized by hot stars rather than AGNs. 
The three objects all show NLS1-like broad emission lines. Castello-Mor et al. (2012) recently
argued that a NLS1-core can be identified in a large fraction of X-ray luminous starforming galaxies.

As an additional test for the driver of the aforementioned correlations, we
estimate $\lambda_{\mathrm{Edd}}=L/L_{\mathrm{Edd}}$ for the final sample by using the parameter 
$L_{\mathrm{[OIII]}}/\sigma_*^4$ as a proxy (e.g., Kauffmann \& Heckman 2006), 
where $L_{\mathrm{[OIII]}}$ and $\sigma_*$ are the [\ion{O}{3}]$\lambda5007$ 
line luminosity and the bulge velocity dispersion, respectively.
The Seyfert 1.5 galaxies are excluded from the estimation because of the 
non-available starlight component in their optical spectra. 
The bolometric luminosity $L$ is transformed from $L_{\mathrm{[OIII]}}$ through the bolometric
correction $L/L_{\mathrm{[OIII]}}\approx3500$ (Heckman et al. 2004).
The [\ion{O}{3}] contribution from star formation is not corrected both because of 
the tight correlation between $L(\mathrm{[OIII]})$ and $L_{\mathrm{2-10kev}}$ and because 
a majority of the final sample is located far from the demarcation lines shown in the BPT diagram (Figure 2). 
The viral blackhole mass $M_{\mathrm{BH}}$ is estimated from the $M_{\mathrm{BH}}-\sigma_*$ relationship:
$\log(M_{\mathrm{BH}}/M_\odot)=8.13+4.02\log(\sigma_*/200\ \mathrm{km\ s^{-1}})$ (Tremaine
et al. 2002), by 1) assuming each galaxy listed in the final sample has a bulge, and 2) using
the velocity dispersion that is measured for each object in the PCA fittings
through the cross-correlation method as a proxy of the value of the bulge.   
In order to obtain the intrinsic velocity dispersion, the instrumental 
resolution is corrected through the equation $\sigma_*^2=\sigma_{\mathrm{obs}}^2-\sigma_{\mathrm{inst}}^2$
by assuming a pure Gaussian profile. The galaxies with
$\sigma_*<60\ \mathrm{km\ s^{-1}}$ (corresponding $\log(M_{\mathrm{BH}}/M_\odot)<6.2$) 
are removed from the following analysis because the
SDSS instrumental resolution is $\sigma_{\mathrm{inst}}\approx60-70\ \mathrm{km\ s^{-1}}$.

The left panel in Figure 7 reproduces the previously reported anti-correlation between $\lambda_{\mathrm{Edd}}$ and the 
$D_n(4000)$ index (e.g., Kewley et al. 2006; Wild et al. 2007; Kauffmann et al. 2007). 
Because the sample size is less than 30, the Kendall's generalized $\tau$ is calculated to
quantify the significance of the correlation. The test returns a Kendall's $\tau=-0.23$ 
($Z=1.721$), and a probability of $P=8.5\times10^{-2}$, where $P$ is the probability that
there is no correlation between the two variables.  $\lambda_{\mathrm{Edd}}$ is plotted against 
$\Gamma_{\mathrm{2-10keV}}$ in the right panel in Figure 7. It is unfortunately that the same statistical
test returns a non-correlation between the two variables ($\tau=-0.02$ and $P=0.859$). 
The negative result might be caused by the following two facts: the small sample size and the
large scatter when $L_{\mathrm{[OIII]}}/\sigma_*^4$ is used as a proxy of $L/L_{\mathrm{Edd}}$ (e.g, Kauffmann et al. 2003).

\subsection{Contamination in Hard, Flat X-ray Spectra}
In addition to the intrinsically low accretion rate, the hard, flat X-ray spectra could be resulted from the other
reasons. Observations show that ``normal'' galaxies are X-ray emitters with a typical X-ray luminosity of 
$10^{38-42}\ \mathrm{erg\ s^{-1}}$ (e,g, Fabbiano 1989; Georgakakis et al. 2008; Brandt et al. 2001). 
The X-ray emission in the ``normal galaxies'' is dominantly emitted from the diffuse 
hot gas ($T\sim10^7$K) and evolved stellar point sources with $T\sim10^6$K (e.g, Fabbiano \& Shaply 2002; Read \& Ponman 2002, and see a 
review in Fabbiano 2006). 
In principle, the X-ray spectra extracted by us contain the emission from 
not only the AGNs, but also the host galaxies, both because of the 
used large aperture and because of the poor spatial resolution.
In order to examine the contamination caused by the host galaxies, 
$\Gamma_{\mathrm{2-10\mathrm{keV}}}$ is plotted 
against the luminosity ratio between AGNs and their host galaxies in the left panel of Figure 8 by
using the [\ion{O}{3}] line luminosity as a proxy of the bolometric luminosity of AGNs. 
The absolute magnitudes of the host galaxies are taken from the catalog provided in Simard et al. (2011) who performed two-dimensional bulge+disk decompositions
for a sample of $\approx10^7$ SDSS-DR7 galaxies\footnote{It is noted that the contamination in the 
absolute magnitude caused by central ANGs is ignorable because our sample is 
dominated by type II AGNs in which the central AGN's continuum is heavily obscured by the torus due to the orientation effect. 
The total light from the host galaxies is adopted because the largest host galaxy half-light radius in our sample is $\approx10\arcsec$, which is
much smaller than the smallest aperture ($=25\arcsec$) used in our X-ray spectral reduction.}. 
The measurements in the $g-$ and $r-$bands are presented by the 
blue-solid and red-open points, respectively.  Two positive correlations 
are expected if there is strong contamination from the host galaxies.
Spearman rank-order tests, however, return two marginal anti-correlations:  
$\rho=-0.221,P=0.105$ for the $g-$band, and $\rho=-0.178,P=0.190$ for the $r-$band. These statistical results 
indicate that the hard, flat X-ray spectra are not caused by the contamination from the host galaxy X-ray emission.

The hard, flat X-ray spectra could be alternatively
caused by the strong Doppler-boosted emission from the relativistic jets.
The average X-ray photon index of radio-loud QSOs is 
about 1.6 (e.g., Reeves \& Turner 2000). By cross-matching the final sample with the FIRST survey catalog (Becker et al. 2003),
$\Gamma_{\mathrm{2-10\mathrm{keV}}}$ is plotted against 
the flux ratio $f_{\mathrm{1.4GHz}}/f_{\mathrm{[OIII]}}$ in the right panel of Figure 8 to examine the 
jet beaming contamination, where $f_{\mathrm{[OIII]}}$ and $f_{\mathrm{1.4GHz}}$ are the [\ion{O}{3}]
line flux and radio flux at 1.4GHz, respectively. The figure shows a positive correlation between the two variables
(with a Kendall's $\tau=0.573$ and $P=0.0498$). Briefly, the stronger the radio emission, the softer and steeper the 
X-ray spectrum will be. The correlation is, however, contrary to the expectation that is inferred from the 
jet beaming hypothesis. The contrary hence suggests that the correlations involving  $\Gamma_{\mathrm{2-10\mathrm{keV}}}$ is
not driven by the contamination from the jet emission.

\subsection{Contamination From Scattered AGN Continuum}

One problem in stellar population synthesis of strong type II AGNs is how to assess the underlying
scattered continuum from central engine (e.g., Cid Fernandes \& Terlevich 1995; Storchi-Bergmann et al. 2000; 
Cid Fernandes et al. 2004), which is implied by the commonly accepted Unified Model (e.g., Antonucci 1993).
The scattered continuum can contaminate $D_n(4000)$ measurement by impacting the spectral shape at the blue end. 
Basically, an underestimated $D_n(4000)$ is expected if the contamination is ignored.      
The difficulty in separating the underlying continuum from observed spectrum mainly comes from the 
strong degeneracy between the continuum and the spectra of young O and B stars (e.g., Cid Fernandes et al. 2004; Hao et al. 2005).  

To address the contamination issue, we formally re-model the continuum of the objects by a linear combination of the used eigenspectra 
and a powerlaw continuum with different index. Again, it is emphasized that one needs to bear in mind the
degeneracy between AGN's continuum and blue spectra of massive stars, which could result in an overestimated of the amplitude of 
the powerlaw (e.g., Cid Fernands et al. 2004), and the same of the inferred $D_n(4000)$. In Figure 9,  
the measured values of $D_n(4000)$ in which the contribution from AGN's continuum is formally corrected are 
compared with that shown in Table 1. The powerlaw index $\lambda$ varies from 1.5 to 1.9. 

\section{SUMMARY}
The evolutionary role of hard X-ray emission is examined by performing 
X-ray and optical spectral analysis on 67 (partially) obscured AGNs selected from the 
\it XMM-Newton\rm\ 2XMMi/SDSS-DR7 catalog. With much better determined 
X-ray photon index, $\Gamma_{\mathrm{2-10keV}}$ is
found to correlate with the stellar population, which is likely 
related to the evolution of $L/L_{\mathrm{Edd}}$. The better determination also allows us to identify two improved 
correlations between $\Gamma_{\mathrm{2-10keV}}$ and the narrow-line ratios 
(i.e., [\ion{O}{1}]/H$\alpha$ and [\ion{S}{2}]/H$\alpha$). Finally, the current sample confirms
the two previously established tight correlations between the two  line ratios and $D_n(4000)$.

\acknowledgments
We thank an anonymous referee for his/her careful review and useful suggestions in improving the manuscript.
The authors thank Profs. Todd A. Boroson and Richard F. Green for providing the optical \ion{Fe}{2} complex
template. This study uses the
SDSS archive data that was created and distributed by the Alfred P. Sloan Foundation.
This work is based on observations obtained with XMM-Newton, an 
ESA science mission with instruments and contributions directly funded by ESA Member States and the USA (NASA).
The study is supported by the National Basic Research Program of China (grant
2009CB824800) and by National Natural Science
Foundation of China under grant 11003022.

\clearpage




\clearpage

\begin{figure}
\includegraphics[scale=0.7]{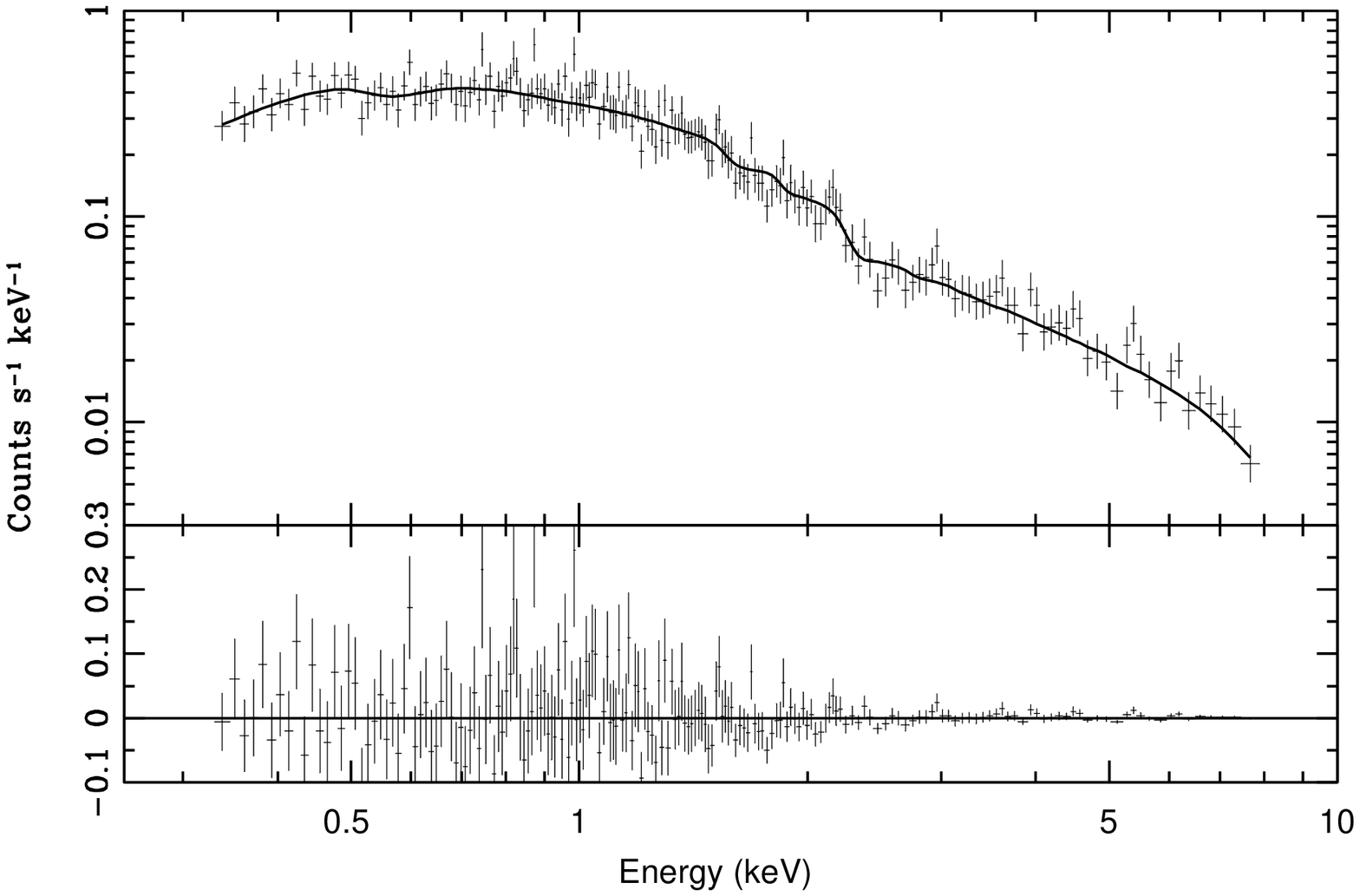}
\caption{\it Top panel:\rm\ EPIC PN X-ray spectrum of SDSS\,J151640.21+001501.8 and the best-fit spectral model 
consisting of an absorbed power-law. \it Bottom panel:\rm\ Deviations, in
unit of $\mathrm{counts\ s^{-1}\ keV^{-1}}$, of the observed data from the best-fit model.}
\end{figure}

\begin{figure}
\epsscale{.80}
\plotone{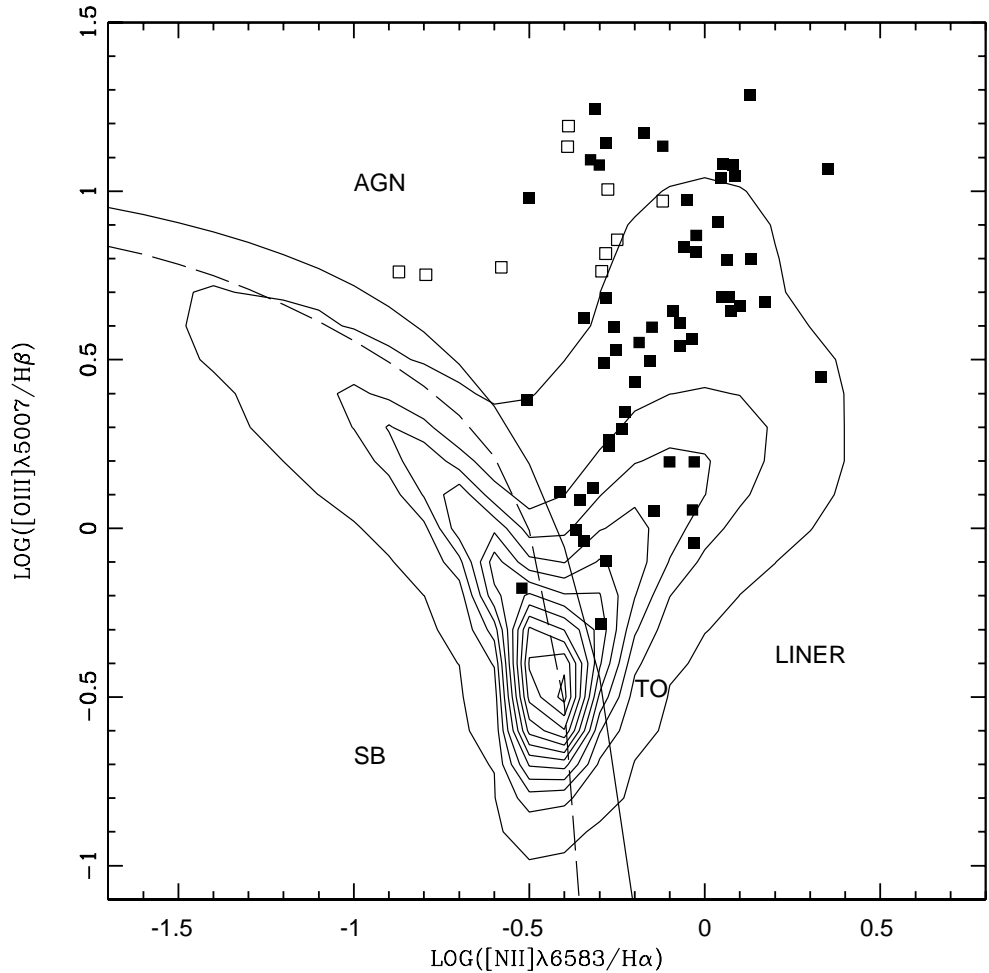}
\caption{Locations on the [\ion{N}{2}]/H$\alpha$ versus [\ion{O}{3}]/H$\beta$ BPT diagnostic diagram for the final sample. 
The narrow emission-line galaxies and broad-line AGNs are shown by the filled and open squares, respectively. 
The broad emission lines are excluded from the calculation of the line ratios. 
The solid line shows the empirical demarcation line proposed by Kauffmann et al. (2003), and the dashed line the 
theoretical line given in Stasinska et al. (2006). Both lines are used to separate `pure' starforming galaxies.
The density contours show a typical
distribution of narrow-line galaxies. Only the galaxies with S/N$>$20 and the emission lines detected with at 
least 3$\sigma$ significance are plotted (Kauffmann et al. 2003). }
\end{figure}

\begin{figure}
\epsscale{.80}
\plotone{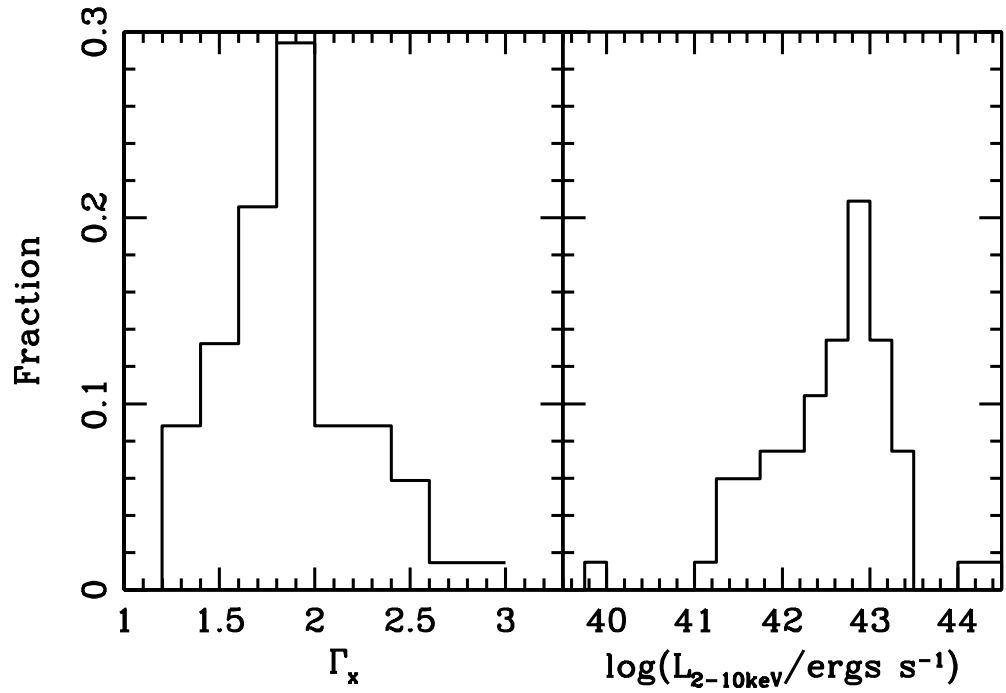}
\caption{\it Left panel:\rm\ Distribution of the X-ray spectral photon index for the final sample. \it Right panel:\rm\
The same as the left one but for the 2-10keV luminosity.}
\end{figure}

\begin{figure}
\epsscale{.80}
\plotone{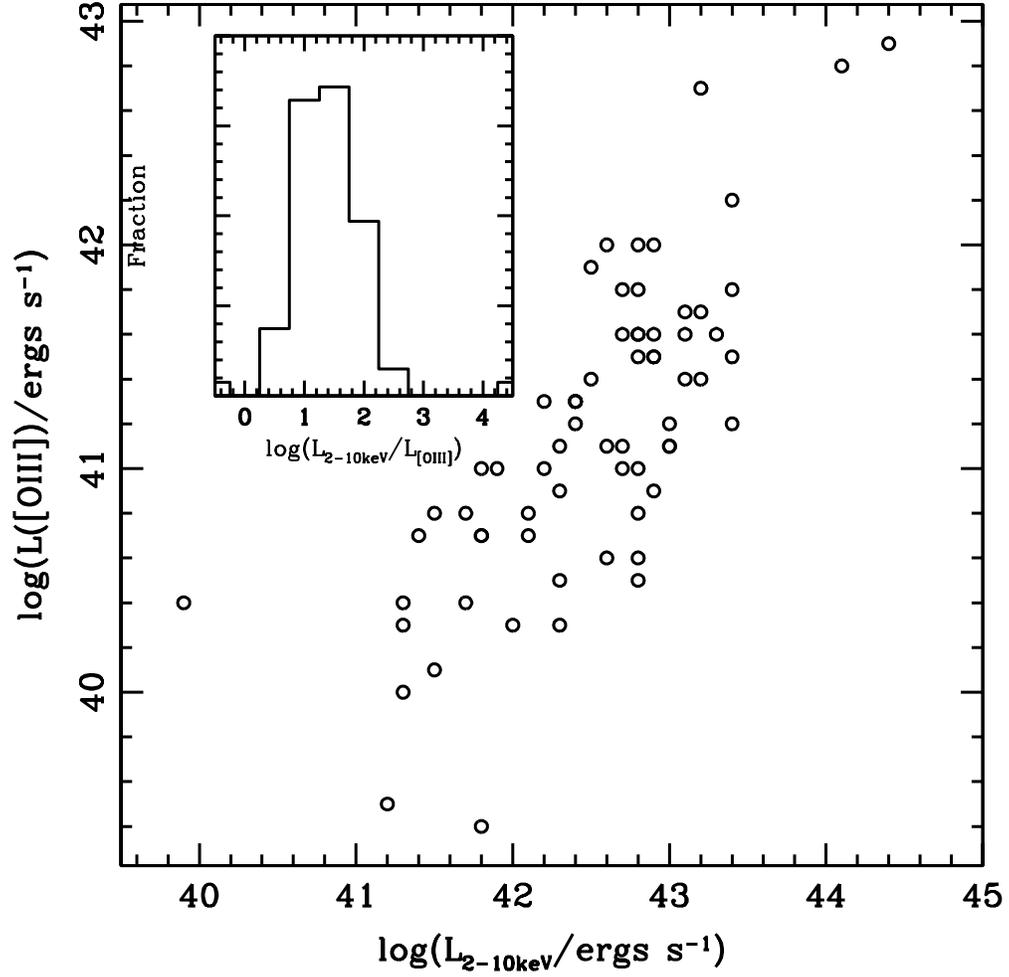}
\caption{Luminosity in the energy bandpass 2-10 keV plotted against the [\ion{O}{3}] luminosity for the final sample. 
The insert panel shows the distribution of the luminosity ratio $L_{\mathrm{2-10keV}}/L_{\mathrm{[OIII]}}$.}
\end{figure}

\begin{figure}
\epsscale{.80}
\plotone{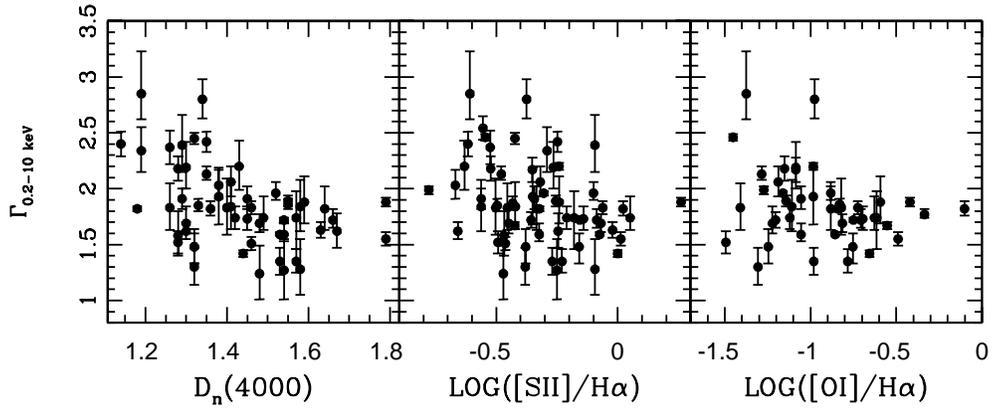}
\caption{$\Gamma_{\mathrm{2-10keV}}$ is plotted against $D_n(4000)$ (left panel),the line ratios 
[\ion{S}{2}]/H$\alpha$ (middle panel) and [\ion{O}{1}]/H$\alpha$ (right panel). The over-plotted 
errorbars correspond to a 1$\sigma$ significance level. The results
from the Spearman rank-order tests are shown in Table 3 for all the three correlations.}
\end{figure}

\begin{figure}
\epsscale{.80}
\plotone{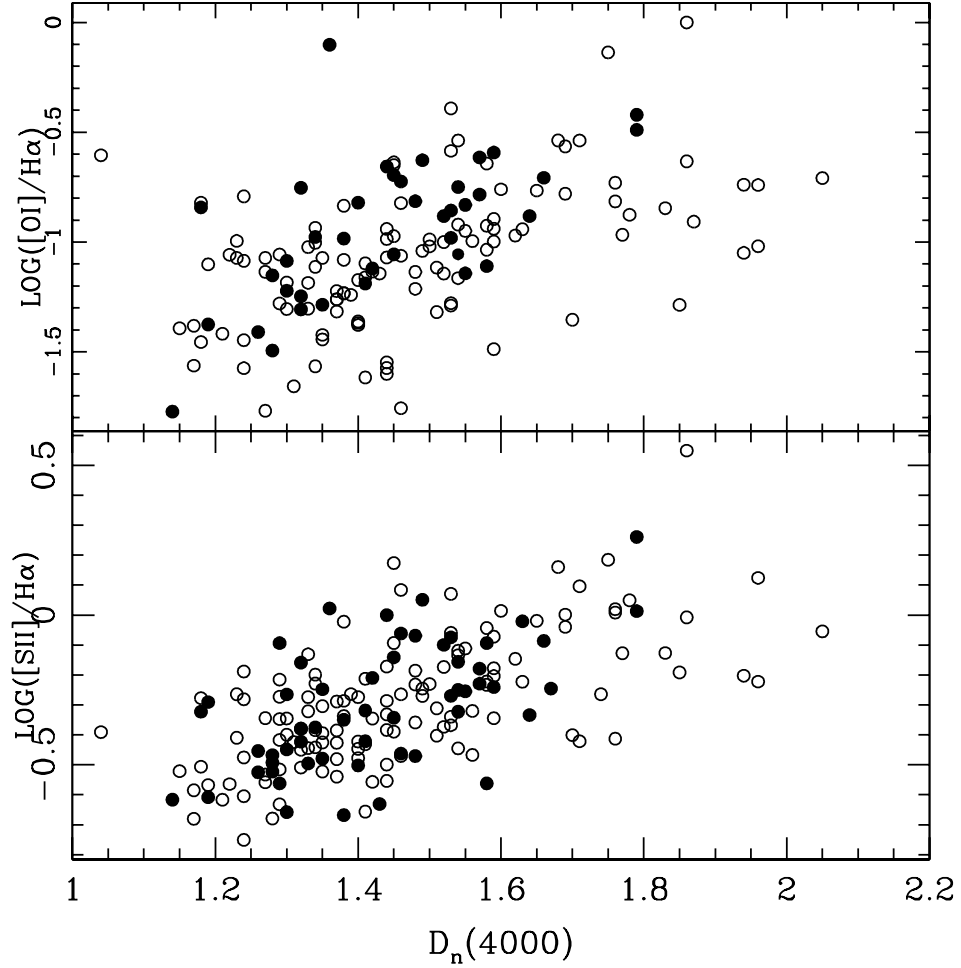}
\caption{\it Upper panel:\rm\ Correlation between the line ratio [\ion{O}{1}]/H$\alpha$ and $D_n(4000)$ for the 
\it XMM-Newton\rm-selected AGNs (the sub-sample). The objects listed in the final sample are marked by the filled circles.
\it Lower panel:\rm\ The same as the upper one but for the line ratio [\ion{S}{2}]/H$\alpha$. The results
from the Spearman rank-order tests are shown in Table 2 for all the correlations. }
\end{figure}

\begin{figure}
\epsscale{.80}
\plotone{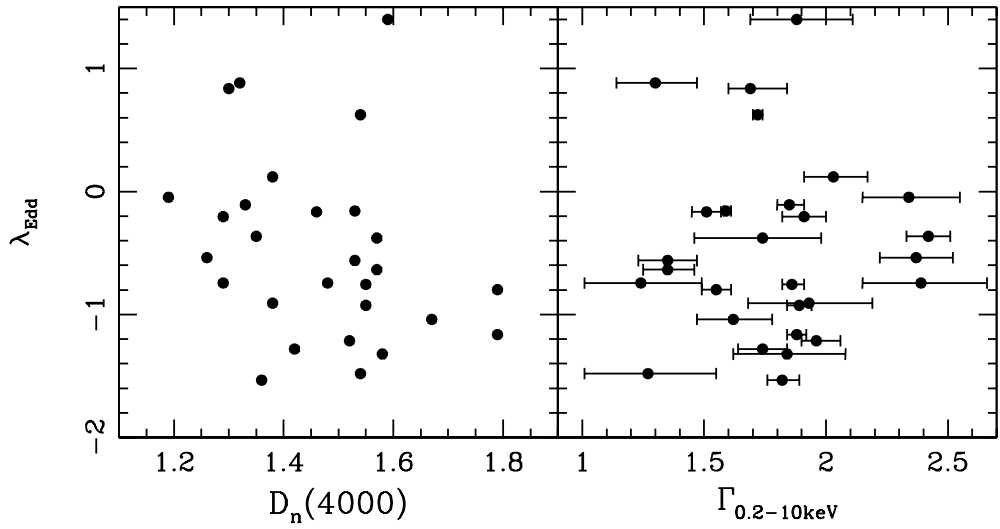}
\caption{$L_{\mathrm{[OIII]}}/\sigma_*^4$ is plotted as a function of 
$D_n(4000)$ (left panel) and $\Gamma_{\mathrm{2-10keV}}$ (right panel). Again, the over-plotted 
errorbars correspond to a 1$\sigma$ significance level.}
\end{figure}
\clearpage

\begin{figure}
\epsscale{.80}
\plotone{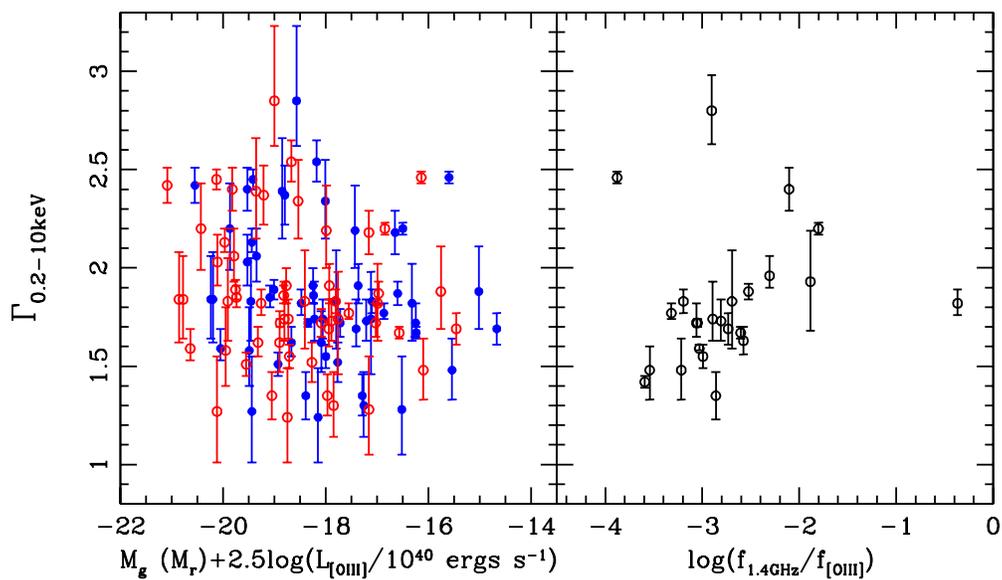}
\caption{\it Left panel:\rm\ $\Gamma_{\mathrm{2-10keV}}$ is plotted against the luminosity ratio
between AGN and its host galaxy. The measurements in the $g-$ and $r-$bands are shown by the blue-solid and red-open 
circles, respectively. \it Right panel:\rm\ $\Gamma_{\mathrm{2-10keV}}$ plotted against the 
radio-to-[\ion{O}{3}] flux ratio.}
\end{figure}
\clearpage

\begin{figure}
\epsscale{.80}
\plotone{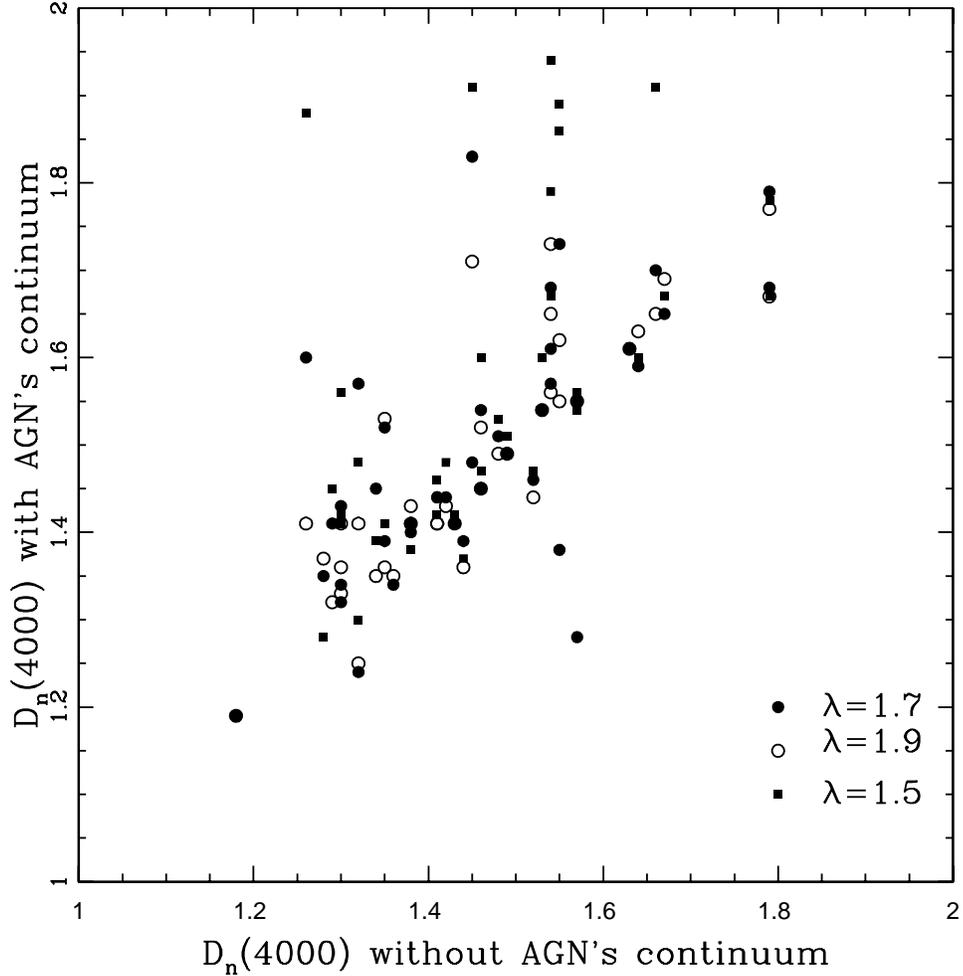}
\caption{The derived values of $D_n(4000)$ in which the contamination of AGN's continuum are formally corrected are 
compared with those shown in Table 1. One must bear in mind that there is a degeneracy between 
AGN's continuum and spectra of massive stars (see the main text for the details). 
The powerlaw index $\lambda$ is fixed to be 1.5, 1.7 and 1.9.}
\end{figure}


\begin{deluxetable}{cccccccccc}
\tabletypesize{\tiny}
\tablewidth{0pt}
\tablecaption{List of Properties of the XMM-Newton/SDSS-DR7 (Partially) Obscured AGNs}
\tablehead{
\colhead{SDSS Name} & \colhead{$z$}  & \colhead{$\mathrm{[NII]/H\alpha}$} & \colhead{$\mathrm{[SII]/H\alpha}$}  &
\colhead{$\mathrm{[OI]/H\alpha}$}    & \colhead{$\mathrm{[OIII]/H\beta}$} & \colhead{$D\mathrm{_n(4000)}$}   &
\colhead{$\Gamma_{\mathrm{2-10keV}}$} & \colhead{$\log L_{\mathrm{2-10keV}}$} & \colhead{$\log L_{\mathrm{[OIII]}}$}\\
\colhead{(1)} & \colhead{(2)} & \colhead{(3)} & \colhead{(4)} & \colhead{(5)} & \colhead{(6)} &
\colhead{(7)} & \colhead{(8)} & \colhead{(9)} & \colhead{(10)} }

\startdata
J000441.24+000711.2  &  0.108  &  -0.069  &  -0.255  &  -1.143  & 0.608  &  1.55  &   $1.89^{+0.05}_{-0.05}$ &  42.8  &  40.6\\
J004311.60$-$093816.0  &  0.054  &  -0.145  &  -0.422  &  \dotfill  & 0.052  &  1.41  &   $1.84^{+0.22}_{-0.20}$ &  42.3  &  40.5\\
J020029.06+002846.6  &  0.174  &  0.100  &  -0.094  &  \dotfill  & 0.658  &  1.29  &   $2.39^{+0.27}_{-0.24}$ &  42.8  &  41.0\\
J024912.86$-$081525.7  &  0.029  &  -0.314  &  -0.343  &  -1.056  & 1.244  &  1.45  &   $1.91^{+0.11}_{-0.09}$ &  41.7  &  40.8\\
J032525.35$-$060837.9  &  0.034  &  -0.058  &  -0.323  &  -0.842  & 0.836  &  1.18  &   $1.82^{+0.02}_{-0.02}$ &  42.5  &  41.9\\
J082433.33+380013.1  &  0.103  &  -0.521  &  -0.617  &  -1.772  & -0.178  &  1.14  &   $2.40^{+0.11}_{-0.11}$ &  42.2  &  41.0\\
J082510.23+375919.6  &  0.021  &  0.128  &  -0.061  &  -0.724  & 1.285  &  1.46  &  $1.83^{+0.06}_{-0.06}$ &  41.9  &  41.0\\
J083737.04+254750.5  &  0.079  &  -0.414  &  -0.502  &  -0.821  & 0.108  &  1.40  &   $1.83^{+0.26}_{-0.24}$ &  43.0  &  41.1\\
J091636.53+301749.3  &  0.123  &  -0.187  &  -0.449  &  -1.222  & 0.551  &  1.30  &   $1.69^{+0.15}_{-0.09}$ &  42.9  &  42.0\\
J091958.02+371128.5  &  0.007  &  0.076  &  -0.021  &  \dotfill  & 0.646  &  1.63  &   $1.63^{+0.07}_{-0.07}$ &  41.8  &  39.4\\
J093952.76+355358.9  &  0.137  &  -0.172  &  -0.241  &  -0.593  & 1.173  &  1.59  &   $1.88^{+0.23}_{-0.19}$ &  43.2  &  42.7\\
J095848.67+025243.2  &  0.079  &  -0.228  &  -0.562  &  -1.110  & 0.344  &  1.58  &   $1.84^{+0.24}_{-0.22}$ &  41.5  &  40.1\\
J095858.53+021459.1  &  0.132  &  -0.090  &  -0.141  &  -0.695  & 0.644  &  1.45  &   $1.73^{+0.11}_{-0.10}$ &  42.9  &  41.6\\
J100035.47+052428.5  &  0.079  &  -0.871  &  -0.778  &  -1.273  & 0.760  &  \dotfill  &   $1.99^{+0.03}_{-0.04}$ &  43.1  &  41.7\\
J101733.20$-$000145.2  &  0.061  &  0.062  &  -0.087  &  -0.707  & 0.795  &  1.66  &   $1.72^{+0.10}_{-0.07}$ &  43.4  &  41.5\\
J101830.79+000504.9  &  0.062  &  -0.287  &  -0.245  &  \dotfill  & 0.490  &  1.67  &   $1.62^{+0.16}_{-0.15}$ &  42.8  &  40.5\\
J101843.13+413516.1  &  0.084  &  0.068  &  0.051  &  -0.628  & 0.687  &  1.49  &   $1.74^{+0.19}_{-0.11}$ &  42.2  &  41.3\\
J102147.85+131228.1  &  0.085  &  0.037  &  -0.156  &  -0.750  & 0.907  &  1.54  &   $1.72^{+0.02}_{-0.02}$ &  43.3  &  41.6\\
J102217.95+212642.8  &  0.042  &  -0.300  &  -0.498  &  -1.494  & -0.283  &  1.28  &  $1.52^{+0.10}_{-0.10}$ &  41.4  &  40.7\\
J102348.44+040553.7  &  0.099  &  -0.030  &  -0.248  &  \dotfill  & 0.197  &  1.35  &   $2.42^{+0.09}_{-0.09}$ &  42.0  &  40.3\\
J102822.84+235125.7  &  0.173  &  -0.294  &  -0.351  &  -1.086  & 0.762  &  \dotfill  &   $2.17^{+0.11}_{-0.11}$ &  43.4  &  42.2\\
J103059.09+310255.7  &  0.178  &  -0.389  &  -0.242  &  -0.984  & 1.193  &  \dotfill  &   $2.20^{+0.03}_{-0.03}$ &  47.2  &  42.7\\
J103701.36+414946.2  &  0.123  &  -0.273  &  -0.631  &  \dotfill  & 0.244  &  1.43  &   $2.20^{+0.23}_{-0.21}$ &  42.1  &  40.7\\
J104930.92+225752.3  &  0.033  &  -0.120  &  -0.179  &  -0.614  & 1.133  &  1.57  &   $1.74^{+0.24}_{-0.28}$ &  43.2  &  41.4\\
J105144.24+353930.7  &  0.159  &  0.079  &  -0.069  &  -0.814  & 1.079  &  1.48  &   $1.69^{+0.08}_{-0.08}$ &  44.1  &  42.8\\
J110101.77+110248.9  &  0.036  &  -0.120  &  \dotfill  &  -0.336  & 0.971  &  \dotfill  &   $1.77^{+0.05}_{-0.03}$ &  43.1  &  41.4\\
J110444.64+381058.9  &  0.143  &  -0.282  &  -0.454  &  -1.410  & -0.098  &  1.26  &  $1.83^{+0.22}_{-0.20}$ &  42.7  &  41.0\\
J111443.66+525834.3  &  0.079  &  -0.507  &  -0.608  &  -1.375  & 0.380  &  1.19  &   $2.85^{+0.38}_{-0.23}$ &  42.3  &  40.9\\
J111552.33+424330.3  &  0.101  &  -0.155  &  -0.463  &  \dotfill  & 0.497  &  1.46  &   $1.51^{+0.06}_{-0.06}$ &  42.3  &  41.1\\
J113409.01+491516.3  &  0.037  &  -0.281  &  -0.334  &  -0.882  & 1.142  &  1.64  &   $1.82^{+0.20}_{-0.19}$ &  42.4  &  41.2\\
J114612.17+202329.9  &  0.023  &  0.331  &  0.261  &  -0.420  & 0.448  &  1.79  &   $1.88^{+0.04}_{-0.04}$ &  41.8  &  40.7\\
J120057.93+064823.1  &  0.036  &  -0.151  &  -0.270  &  -0.981  & 0.596  &  1.53  &  $1.35^{+0.12}_{-0.12}$ &  43.4  &  41.2\\
J120442.10+275411.7  &  0.165  &  -0.391  &  -0.305  &  -1.161  & 1.132  &  \dotfill  &   $1.96^{+0.02}_{-0.02}$ &  44.4  &  42.9\\
J121049.60+392822.1  &  0.023  &  -0.035  &  -0.100  &  -0.882  & 0.054  &  1.52  &   $1.96^{+0.10}_{-0.06}$ &  41.3  &  40.0\\
J121118.86+503652.6  &  0.102  &  -0.281  &  -0.381  &  -1.307  & 0.684  &  1.32  &   $1.30^{+0.17}_{-0.16}$ &  43.4  &  41.8\\
J122137.93+043026.1  &  0.095  &  -0.276  &  -0.432  &  \dotfill  & 1.005  &  \dotfill  &   $1.87^{+0.06}_{-0.06}$ &  42.8  &  41.8\\
J122546.72+123942.7  &  0.009  &  -0.326  &  -0.159  &  -0.753  & 1.093  &  1.32  &   $1.48^{+0.12}_{-0.15}$ &  42.9  &  41.5\\
J122649.56+311736.3  &  0.083  &  -0.026  &  -0.229  &  -0.783  & 0.820  &  1.57  &   $1.35^{+0.11}_{-0.10}$ &  42.9  &  41.5\\
J122934.04+134629.3  &  0.099  &  0.046  &  -0.265  &  -1.086  & 1.039  &  1.30  &   $2.19^{+0.23}_{-0.19}$ &  42.4  &  41.3\\
J124512.93$-$004056.5  &  0.104  &  -0.346  &  -0.668  &  \dotfill  & -0.036  &  1.38  &   $2.03^{+0.14}_{-0.12}$ &  41.8  &  40.7\\
J124828.44+083112.7  &  0.119  &  -0.025  &  \dotfill  &  -0.831  & 0.868  &  1.55  &   $1.86^{+0.05}_{-0.04}$ &  43.0  &  41.1\\
J130845.69$-$013053.9  &  0.111  &  -0.502  &  -0.523  &  -1.152  & 0.981  &  1.28  &   $2.18^{+0.11}_{-0.11}$ &  42.7  &  41.6\\
J132525.63+073607.5  &  0.124  &  0.053  &  -0.094  &  \dotfill  & 1.079  &  1.58  &   $1.28^{+0.27}_{-0.23}$ &  42.7  &  41.8\\
J132925.36+115754.9  &  0.148  &  -0.275  &  -0.525  &  \dotfill  & 0.262  &  1.26  &   $2.37^{+0.15}_{-0.15}$ &  42.4  &  41.3\\
J133217.89+291320.0  &  0.136  &  -0.356  &  -0.468  &  \dotfill  & 0.085  &  1.28  &   $1.58^{+0.22}_{-0.18}$ &  42.1  &  40.8\\
J133739.87+390916.4  &  0.020  &  0.351  &  -0.375  &  -0.977  & 1.066  &  1.34  &   $2.80^{+0.18}_{-0.17}$ &  39.9  &  40.4\\
J134054.36+262253.8  &  0.187  &  -0.036  &  -0.291  &  \dotfill  & 0.562  &  1.19  &   $2.34^{+0.21}_{-0.19}$ &  42.8  &  41.6\\
J134208.38+353915.4  &  0.003  &  0.084  &  0.000  &  -0.656  & 1.045  &  1.44  &   $1.42^{+0.03}_{-0.03}$ &  41.2  &  39.5\\
J134351.06+000434.7  &  0.074  &  -0.580  &  -0.554  &  \dotfill  & 0.774  &  \dotfill  &   $2.54^{+0.11}_{-0.10}$ &  41.8  &  41.0\\
J134834.94+263109.8  &  0.059  &  -0.283  &  -0.545  &  -1.452  & 0.815  &  \dotfill  &   $2.46^{+0.03}_{-0.03}$ &  42.6  &  42.0\\
J135553.52+383428.7  &  0.050  &  -0.795  &  -0.423  &  -0.553  & 0.752  &  \dotfill  &   $1.67^{+0.03}_{-0.03}$ &  43.2  &  41.7\\
J140515.58+542458.0  &  0.083  &  0.171  &  -0.250  &  \dotfill  & 0.672  &  1.54  &   $1.27^{+0.28}_{-0.26}$ &  42.3  &  40.3\\
J141314.87$-$031227.3  &  0.006  &  -0.054  &  -0.074  &  -0.857  & 0.973  &  1.53  &   $1.59^{+0.02}_{-0.01}$ &  42.8  &  41.6\\
J142904.60+012017.3  &  0.101  &  -0.319  &  -0.658  &  \dotfill  & 0.121  &  1.30  &   $1.62^{+0.08}_{-0.07}$ &  42.9  &  40.9\\
J143450.62+033842.5  &  0.028  &  -0.344  &  -0.479  &  -1.286  & 0.623  &  1.35  &   $2.13^{+0.07}_{-0.05}$ &  41.3  &  40.3\\
J145442.23+182937.1  &  0.116  &  -0.235  &  -0.471  &  \dotfill  & 0.295  &  1.48  &   $1.24^{+0.25}_{-0.23}$ &  42.8  &  41.5\\
J150121.14+013813.4  &  0.035  &  -0.032  &  -0.350  &  -0.984  & -0.045  &  1.38  &   $1.93^{+0.26}_{-0.25}$ &  41.5  &  40.8\\
J151106.41+054122.9  &  0.081  &  -0.301  &  -0.379  &  -1.247  & 1.077  &  1.32  &   $1.48^{+0.16}_{-0.15}$ &  42.8  &  42.0\\
J151616.85+000804.4  &  0.092  &  -0.070  &  -0.210  &  -1.120  & 0.542  &  1.42  &   $1.74^{+0.10}_{-0.10}$ &  42.6  &  41.1\\
J151640.21+001501.9  &  0.052  &  0.048  &  0.022  &  -0.102  & 0.685  &  1.36  &   $1.82^{+0.07}_{-0.06}$ &  43.0  &  41.2\\
J151741.72+424820.1  &  0.116  &  -0.366  &  -0.562  &  \dotfill  & -0.004  &  1.29  &   $1.91^{+0.09}_{-0.09}$ &  42.7  &  41.1\\
J152203.33+275100.9  &  0.075  &  -0.258  &  -0.496  &  \dotfill  & 0.596  &  1.33  &   $1.85^{+0.06}_{-0.05}$ &  42.6  &  40.6\\
J153152.28+241429.7  &  0.096  &  -0.100  &  -0.423  &  \dotfill  & 0.197  &  1.32  &   $2.45^{+0.05}_{-0.05}$ &  42.8  &  40.8\\
J155855.79+024833.9  &  0.046  &  0.132  &  0.013  &  -0.450  & 0.799  &  1.79  &   $1.55^{+0.06}_{-0.06}$ &  42.5  &  41.4\\
J160452.45+240241.6  &  0.088  &  -0.249  &  -0.357  &  -1.204  & 0.856  &  \dotfill  &  $1.72^{+0.07}_{-0.07}$ &  43.1  &  41.6\\
J160534.64+323940.8  &  0.030  &  -0.254  &  -0.319  &  -1.188  & 0.530  &  1.41  &   $2.06^{+0.14}_{-0.13}$ &  41.3  &  40.4\\
J231815.66+001540.1  &  0.030  &  -0.199  &  -0.323  &  -1.055  & 0.434  &  1.54  &   $1.59^{+0.10}_{-0.06}$ &  41.7  &  40.4\\ 
\enddata
\end{deluxetable}

\begin{table}
\begin{center}
\caption{Spearman Rank-order Test Matrix for the X-ray Spectral Photon Index $\Gamma_{\mathrm{2-10keV}}$\label{tbl-3}}
\begin{tabular}{cccc}
\tableline\tableline
 & $D_n(4000)$ & [\ion{O}{1}]/H$\alpha$ & [\ion{S}{2}]/H$\alpha$\\
\tableline
$\rho$ & -0.382 & -0.332 & -0.329\\
$P$    & $4.2\times10^{-3}$ & $7.9\times10^{-3}$ & $2.6\times10^{-2}$\\
\tableline
\end{tabular}
\end{center}
\end{table}

\begin{table}
\begin{center}
\caption{Spearman Rank-order Correlation Coefficients for the Correlations Between $D_n(4000)$ and the Two 
Line Ratios.\label{tbl-2}}
\begin{tabular}{lcc}
\tableline\tableline
Sample & [\ion{O}{1}]/H$\alpha$ & [\ion{S}{2}]/H$\alpha$\\
\tableline
sub-sample & 0.537($<10^{-4}$) & 0.622($<10^{-4}$)\\
Final sample\dotfill & 0.659($1\times10^{-4}$) & 0.562($<10^{-4}$)\\
\tableline
\end{tabular}
\end{center}
\end{table}


\begin{thebibliography}{}


\bibitem[Abazajian et al. 2009]{aba09} Abazajian, K. N., Adelman-McCarthy, J. K., Agueros, M. A., et al. 2009, \apjs, 182, 543
\bibitem[Alonso-Herrero et al. 2008]{alo08} Alonso-Herrero, A., Perez-Gonzalez, P. G., Rieke, G. H., Alexander, D. M., Rigby, J. R., Papovich, C., Donley, J. L., 
\& Rigopoulou, D. 2008, \apj, 677, 127
\bibitem[Antonucci 1993]{ant93} Antonucci, R. R. J. 1993, \araa, 31, 473
\bibitem[Aird et al. 2010]{air10} Aird, J., et al. 2010, \mnras, 401, 2531
\bibitem[Arnaud 1996]{arn96} Arnaud, K. A. 1996, in ASP Conf. Ser. 101, Astronomical Data Analysis Software and Systems V, ed. 
G. H. Jacoby and J. Barnes, (San Francisco, CA: ASP), 17
\bibitem[Assef et al. 2011]{ass01} Assef, R. J., et al. 2011, \apj, 728, 56
\bibitem[Balogh et al. 1999]{bal99} Balogh, M. L., et al. 1999, \apj, 527, 54
\bibitem[Baldwin et al. 1981]{bal81}Baldwin, J. A., Phillips, M. M., \& Terlevich, R. 1981, \pasp, 93, 5
\bibitem[Becker et al. 2003]{bec03} Becker, R. H., Helfand, D. J., White, R. L., Gregg, M. D., \& Laurent-Muehleisen, S. A. 2003, 
VizieR Online Data Catalog, 8071, 0
\bibitem[Bian 2005]{bia05} Bian, W. H., 2005, \cjaa, 5, 289
\bibitem[Boller et al. 1996]{bol96} Boller, T., Brandt, W. N., \& Fink, H. 1996, \aap, 305, 53
\bibitem[Bongiorno et al. 2007]{bon07} Bongiorno, A., et al. 2007, \aap, 472, 443
\bibitem[Boroson 2002]{bor02} Boroson, T. A. 2002, \apj, 565, 78
\bibitem[Boroson \& Green 1992]{bog92} Boroson, T. A., \& Green, R. F. 1992, \apjs, 80, 109
\bibitem[Brandt et al. 1997]{bra97} Brandt, W. N., Mathur, S., Reynolds, C. S., \& Elvis, M. 1997, \mnras, 292, 407
\bibitem[Brandt et al. 2001]{bra01} Brandt, W. N., et al. 2001, \aj, 122, 2810 
\bibitem[Bromley et al. 1998]{bro98} Bromley, B. C., Press, W. H., Lin, H., \& Kirshner, R. P. 1998, \apj, 505, 25
\bibitem[Brinchmann et al. 2004]{bri04} Brinchmann, J., Charlot, S., White, S. D. M., Tremonti, C., Kauffmann, G., Heckman, T., \& Brinkmann, J. 2004, \mnras, 351, 1151
\bibitem[Bruzual 1983]{bru83} Bruzual, A. G. 1983, \apj, 273, 105
\bibitem[Bruzual \& Charlot 2003]{bc03}Bruzual, G., \& Charlot, S. 2003, \mnras, 344, 100
\bibitem[Caccianiga et al. 2008]{cac08} Caccianiga,  A. et al. 2008, \aap, 477, 735
\bibitem[Cao 2009]{cao09} Cao, X. 2009, \mnras, 394, 207
\bibitem[Cardelli et al. 1989]{car89} Cardelli, J. A., Clayton, G. C., \& Mathis, J. S. 1989, \apj, 345, 245
\bibitem[Castello-Mor et al. 2012]{cat12} Castello-Mor, N., Barcons, X., Ballo, L., Carrera, F. J., Ward, M. J., \& Jin, C. 2012, \aap, 544, 48
\bibitem[Chen et al. 2009]{che09} Chen, Y.M., Wang, J. M., Yan, C. S., Hu, C., \& Zhang, S. 2009, \apjl, 695, 130
\bibitem[Cid Fernandes \& Terlevich 1995]{cit95} Cid Fernandes, R. Jr., \& Terlevich, R. 1995, \mnras, 272, 423
\bibitem[Cid Fernandes et al. 2004]{cid04} Cid Fernandes, R., Gu, Q., Melnick, J., Terlevich, E., Terlevich, R., Kunth, D., Rodrigues Lacerda, R., \& Joguet, B. 2004, \mnras, 355, 273
\bibitem[Collin \& Kawaguchi 2004]{cok04}Collin, S., \&  Kawaguchi, T. 2004, \aap, 426, 797
\bibitem[Comastri 2000]{com00} Comastri, A. 2000, \nar, 44, 403
\bibitem[Corral et al. 2011]{cor11} Corral, A., Della Ceca, R., Caccianiga, A., Severgnini, P., Brunner, H., Carrera, F. J., Page, M. J., 
\& Schwope, A. D. 2011, \aap, 530, 42
\bibitem[Cowie et al. 2003]{cow03} Cowie, L. L., Barger, A. J., Bautz, M. W., Brandt, W. N., \& Garmire, G. P. 2003, \apjl, 584, 57 
\bibitem[Croom et al. 2004]{cro04} Croom, S. M., Smith, R. J., Boyle, B. J., Shanks, T., Miller, L., Outram, P. J., \& Loaring, N. S. 2004, \mnras, 349, 1379
\bibitem[Dadina 2008]{dad08} Dadina, M. 2008, \aap, 485, 417
\bibitem[Davis et al. 2007]{dav07}Davies, R., Mueller Sanchez, F., Genzel, R., et al. 2007, \apj, 671, 1388
\bibitem[Deo et al. 2006]{deo06}Deo, R. P., Crenshaw, D. M., \& Kraemer, S. B. 2006, \aj, 132, 321
\bibitem[Desroches et al. 2009]{des09} Desroches, L., Greene, J. E., \& Ho, L. C. 2009, \apj, 698, 1515
\bibitem[Fabbiano 1989]{fab89} 	Fabbiano, G. 1989, \araa, 27, 87
\bibitem[Fabbiano 2006]{fab06} 	Fabbiano, G. 2006, \araa, 44, 323
\bibitem[Fabbiano \& Shaply 2002]{fas02} Fabbiano, G., \& Shapley, A. 2002, \apj, 565, 908
\bibitem[Ferrarese \& Ford 2005]{fef05} Ferrarese, L., \&  Ford, H. 2005, \ssr, 116, 523	
\bibitem[Ferrarese \& Merritt 2000]{fem00} Ferrarese, L., \& Merritt, D. 2000, \apjl, 539, 9
\bibitem[Flohic et al. 2006]{flo06} Flohic, H. M. L. G., Eracleous, M., Chartas, G., Shields, J. C., \& Moran, E. C. 2006, \apj, 647, 140
\bibitem[Francis et al. 1992]{fra92} Francis, P. J., Hewett, P. C., Foltz, C. B., \& Chaffee, F. H. 1992, \apj, 398, 476
\bibitem[Gebhardt et al. 2000a]{geb00a} Gebhardt, K. et al. 2000a, \apjl, 539, 13
\bibitem[Gebhardt et al. 2000b]{geb00b} Gebhardt, K. et al. 2000b, \apjl, 543, 5
\bibitem[Georgakakis 2008]{geo08} Georgakakis, A. 2008, AN, 329, 174 
\bibitem[Georgantopoulos \& Akylas 2010]{gea10} Georgantopoulos, I., \& Akylas, A. 2010, \aap, 509, 38
\bibitem[Gierlinski \& Done 2004]{gie04} Gierlinski, M., \& Done, C. 2004, \mnras, 349, L7
\bibitem[Glazebrook et al. 1998]{gal98} Glazebrook, K., Offer, A. R., \& Deeley, K. 1998, \apj, 492, 98
\bibitem[Gliozzi et al. 2008]{gli08} Gliozzi, M., Foschini, L., Sambruna, R. M., \& Tavecchio, F. 2008, \aap, 478, 723
\bibitem[Goulding et al. 2010]{gou10} Goulding, A. D., Alexander, D. M., Lehmer, B. D., \& Mullaney, J. R. 2010, \mnras, 406, 597
\bibitem[Greene \& Ho 2006]{grh06} Greene, J. E., \& Ho, L. C. 2006, \apjl, 641, 21
\bibitem[Greene et al. 2008]{gre08} Greene, J. E., Ho, L. C., \& Barth, A. J. 2008, \apj, 688, 159
\bibitem[Grupe 1996]{gru96} Grupe D. 1996, PhD thesis, Univ. Gottingen
\bibitem[Greenhill et al. 2008]{gre08} Greenhill, L. J., Tilak, A., \& Madejski, G. 2008, \apjl, 686, 13
\bibitem[Grupe 2004]{gru04} Grupe, D. 2004, \aj, 127, 1799
\bibitem[Gu \& Cao 2009]{guc09} Gu, M., \&  Cao, X. 2009, \mnras, 399, 349
\bibitem[Gultekin et al. 2009]{gul09} Gultekin, K., et al. 2009, \apj, 698, 198
\bibitem[Haardt \& Maraschi 1991]{ham91} Haardt, F., \& Maraschi, L. 1991, \apjl, 380, 51 
\bibitem[Hao et al. 2005]{hao05} Hao, L., et al. 2005, \aj, 129, 1795
\bibitem[Haring \& Rix 2004]{har04} Haring, N., \& Rix, H. W. 2004, \apjl, 604, 89
\bibitem[Hasinger et al. 2005]{has05} Hasinger, G., Miyaji, T., \& Schmidt, M. 2005, \aap, 441, 417
\bibitem[Heckman \& Kauffman 2006]{hek06} Heckman, T. M., \& Kauffmann, G. 2006, \nar, 50, 677
\bibitem[Heckman et al. 2004]{hec04} Heckman, T. M., Kauffmann, G., Brinchmann, J., et al. 2004. \apj, 613, 109
\bibitem[Heckman et al. 2005]{hec05} Heckman, T. M., Ptak, A., Hornschemeier, A., \& Kauffmann, G. 2005, \apj, 634, 161
\bibitem[Hopkins et al. 2005]{hop05}Hopkins, P. F., Hernquist, L., Martini, P., et al. 2005, \apj, 625, 71
\bibitem[Imanishi \& Wada 2004]{imw04} Imanishi, M., \& Wada, K. 2004, \apj, 617, 241
\bibitem[Jansen et al. 1999]{jan99} Jansen, F., et al. 2001, \aap, 365, L1
\bibitem[Jin et al. 2012]{jin12} Jin, C., Ward, M., \& Done, C. 2012, \mnras, 425, 907
\bibitem[Just et al. 2007]{jus07} Just, D. W., Brandt, W. N., Shemmer, O., Steffen, A. T., Schneider, D. P., Chartas, G., \& Garmire, G. P. 2007, \apj, 665, 1004
\bibitem[Kalberla et al. 2005]{kal05}Kalberla, P. M. W., Burton, W. B., Hartmann, D., Arnal, E. M., Bajaja, E., Morras, R., \& Poppel, W. G. L. 2005, 
\aap, 440, 775 
\bibitem[Kauffmann \& Heckman 2009]{kah09} Kauffmann, G., \& Heckman, T. M. 2009, \mnras, 397, 135
\bibitem[Kauffmann et al. 2003]{kau03}Kauffmann, G., Heckman, T. M., White, S. D. M., et al. 2003, \mnras, 341, 33
\bibitem[Kauffmann et al. 2007]{kau07} Kauffmann, G., et al. 2007, \apjs, 173, 357
\bibitem[Kawaguchi et al. 2001]{kwa01}Kawaguchi, T., Shimura, T., \& Mineshige, S. 2001, \apj, 546, 966 
\bibitem[Kelly et al. 2008]{kel08} Kelly, B. C., Bechtold, J., Trump, J. R., Vestergaard, M., \& Siemiginowska, A. 2008, \apjs, 176, 355
\bibitem[Kewley et al. 2006]{kew06} Kewley, L. J., Groves, B., Kauffmann, G., \& Heckman, T. 2006, \mnras, 372, 961
\bibitem[Komossa 2008]{kom08} Komossa, S. 2008, RevMexAA Conf. Ser., 32, 86
\bibitem[Kriss 1994]{kri94} Kriss, G. 1994, Adass, 3, 43
\bibitem[LaMassa et al. 2011]{lam11} LaMassa, S. M., Heckman, T. M., Ptak, A., Martins, L., Wild, V., Sonnentrucker, P., \& Hornschemeier, A. 2011, \apj, 729, 52
\bibitem[Laor et al. 1994]{lao94}Laor, A., Fiore, F., Elvis, M., Wilkes, B. J., \& McDowell, J. C. 1994, \apj, 435, 611
\bibitem[Laor et al. 1997]{lao97}Laor, A., Fiore, F., Elvis, M., Wilkes, B. J., \& McDowell, J. C. 1997, \apj, 477, 93 
\bibitem[Leighly 1999]{lei99} Leighly, K. M. 1999, \apjs, 125, 317
\bibitem[Li et al. 2008]{li08} 	Li, C., Kauffmann, G., Heckman, T., M., White, S. D. M., \& Jing, Y. P. 2008, \mnras, 385, 1915 
\bibitem[Li et al. 2005]{li05}  Li, C., Wang, T. G., Zhou, H. Y., Dong, X, B., \& Cheng, F, Z. 2005, \aj, 129, 669
\bibitem[Lira et al. 2002a]{lir02a} Lira, P., Ward, M., Zezas, A., Alonso-Herrero, A., \& Ueno, S. 2002a, \mnras, 330, 259
\bibitem[Lira et al. 2002b]{lir02b} Lira, P., Ward, M., Zezas, A., \& Murray, S. S. 2002b, \mnras, 333, 709
\bibitem[Lu \& Yu 1999]{luy99} Lu, Y. J. \& Yu, Q. J. 1999, \apjl, 526, 5
\bibitem[Magdziarz \& Zdziarski 1995]{maz95} Magdziarz, P., \&  Zdziarski, A. A. 1995, \mnras, 273, 837
\bibitem[Magorrian et al. 1998]{mag98} Magorrian, J., et al. 1998, \aj, 115, 2285
\bibitem[Mao et al. 2009]{mao09} Mao, Y. F., Wang, J., \& Wei, J. Y. 2009, \apj, 698, 859 
\bibitem[Marziani \& Sulentic 2012]{mas12} Marziani, P., \& Sulentic, J. W. 2012, New Astronomy Review, 56, 49
\bibitem[Mateos et al. 2010]{mat10} Mateos, S., et al. 2010, \aap, 510, 35
\bibitem[Mathur 2000]{mat00} Mathur, S. 2000, \mnras, 314, L17
\bibitem[McLure \& Dunlop 2002]{mcd02} McLure, R. J., \& Dunlop, J. S. 2002, \mnras, 331, 795
\bibitem[McLure \& Dunlop 2004]{mcd04} McLure, R. J., \& Dunlop, J. S. 2004, \mnras, 352, 1390
\bibitem[Merritt \& Ferrarese 2001]{mef01} Merritt, D., \& Ferrarese, L. 2001, \mnras, 320, L30
\bibitem[Mushotzky et al. 1980]{nus80} Mushotzky, R. F., Marshall, F. E., Boldt, E. A., Holt, S. S., \& Serlemitsos, P. J. 1980, \apj, 235, 377
\bibitem[Nandra et al. 2005]{nan05} Nandra, K., Laird, E. S., \& Steidel, C. C. 2005, \mnras, 360, L39
\bibitem[Nandra et al. 1990]{nan90} Nandra, K., Pounds, K. A., \& Stewart, G. C. 1990, \mnras, 242, 660
\bibitem[Nandra et al. 2007]{nan07} Nandra, K., et al. 2007, \apjl, 660, 11
\bibitem[O'Sullivan et al. 2001]{osu01} O'Sullivan, E., Forbes, D. A., \& Ponman, T. J. 2001, \mnras, 342, 420
\bibitem[Page et al. 2005]{pag05} Page, K. L., Reeves, J. N., O'Brien, P. T., \& Turner, M. J. L. 2005, \mnras, 364, 195
\bibitem[Panessa et al. 2006]{pan06} Panessa, F., et al. 2006, \aap, 455, 173
\bibitem[Panessa et al. 2008]{pan08} Panessa, F., et al. 2008, \aap, 483, 151
\bibitem[Piconcelli et al. 2005]{pic05} Piconcelli, E., Jimenez-Bailon, E., Guainazzi, M., Schartel, N., Rodriguez-Pascual, P. M., \& Santos-Lleo, M. 2005, \aap, 432, 15
\bibitem[Pineau et al. 2011]{pin11} Pineau, F.-X., Motch, C., Carrera, F., Della Ceca, R., Derriere, S., Michel, L., Schwope, A., \& Watson, M. G.
2011, \aap, 527, 126
\bibitem[Porquet et al. 2004]{por04} Porquet, D., Reeves, J. N., O'Brien, P., \& Brinkmann, W. 2004, \aap, 422, 85
\bibitem[Pounds et al. 1995]{pou95} Pounds, K. A., Done, C., \& Osborne, J. P. 1995, \mnras, 277, L5
\bibitem[Puchnarewicz et al. 1992]{puc92} Puchnarewicz, E. M., et al. 1992, \mnras, 256, 589
\bibitem[Read \& Ponman 2001]{rep01} Read, A. M., \& Ponman, T. J. 2001, \mnras, 328, 127
\bibitem[Reeves \& Turner 2000]{ret00} Reeves, J. N., \& Turner, M. J. L. 2000, \mnras, 316, 243
\bibitem[Reichard et al. 2008]{rei08} Reichard, T. A., Heckman, T. M., Rudnick, G., Brinchmann, J., \& Kauffmann, G. 2008, \apj, 677, 186
\bibitem[Rinn et al. 2005]{rin05} Rinn, A. S., Sambruna, R. M., \& Gliozzi, M. 2005, \apj, 621, 167
\bibitem[Risaliti et  al. 2009]{ris09}	Risaliti, G., Young, M., \& Elvis, M. 2009, \apjl, 700, 6 
\bibitem[Sanchez et al. 2004]{san04} Sanchez, S. F., et al. 2004, \apj, 641, 586
\bibitem[Sani et al. 2010]{san10}Sani, E., Lutz, D., Risaliti, G., Netzer, H., Gallo, L. C., Trakhtenbrot, B., Sturm, E., \& Boller, T. 2010, \mnras, 403, 1246 
\bibitem[Schartmann et al. 2010]{sch10} Schartmann, M., Burkert, A., Krause, M., Camenzind, M., Meisenheimer, K., \& Davies, R. I. 2010,\mnras, 403, 1801
\bibitem[Schawinski et al. 2007]{sch07} Schawinski, K., Thomas, D., Sarzi, M., Maraston, C., Kaviraj, S., Joo, S., Yi, S. K., \& Silk, J. 2007, \mnras, 382, 1415
\bibitem[Schawinski et al. 2009]{sch09} Schawinski, K., Virani, S., Simmons, B., Urry, C. M., Treister, E., Kaviraj, S., \& Kushkuley, B. 2009, \apjl, 692, 19
\bibitem[Shankar et al. 2009]{sha09} Shankar, F., Bernardi, M., \& Haiman, Z. 2009, \apjs, 694, 867
\bibitem[Shemmer et al. 2006]{she06} Shemmer, O., Brandt, W. N., Netzer, H., Maiolino, R., \& Kaspi, S. 2006, \apjl, 646, 29 
\bibitem[Shemmer et al. 2005]{she05} Shemmer, O., Brandt, W. N., Vignali, C., Schneider, D. P., Fan, X., Richards, G. T., \& Strauss, M. A. 2005, \apj, 630, 729
\bibitem[Shemmer et al. 2008]{she08} Shemmer, O., et al. 2008, \apj, 682, 81\bibitem[Schlegel et al. 1998]{sch98} Schlegel, D., Finkbeiner, D. P., \& Davis, M. 1998, \apj, 500, 525
\bibitem[Silverman, et al. 2008]{sil08} Silverman, J. D., et al. 2008, \apjs, 679, 118
\bibitem[Simard et al. 2011]{sim11} Simard, L., Mendel, J. T., Patton, D. R., Ellison, S., L., \& McConnachie, A. W. 2011, \apjs, 196, 11
\bibitem[Singh et al. 2011]{sin11} Singh, V., Shastri1, P., \& Risaliti, G. 2011, \aap, 532, 84
\bibitem[Spergel et al. 2003]{spe03} Spergel, D. N., et al. 2003, \apjs, 148, 175
\bibitem[Storchi-Bergmann et al. 2000]{stb00} Storchi-Bergmann, T., Raimann, D., Bica, E. L. D., \& Fraquelli, H. A. 2000, \apj, 544, 747
\bibitem[Struder et al. 2001]{str01} Struder, L., et al. 2001, \aap, 365, L18
\bibitem[Sulentic et al. 2000]{sul00} Sulentic, J. W., Marziani, P., \& Dultzin-Hacyan, D. 2000, \araa, 38, 521
\bibitem[Tremaine et al. 2002]{tre02} Tremaine, S., Gebhardt, K., Bender, R., et al. 2002, \apj, 574, 740
\bibitem[Treister et al. 2009]{tre09} Treister, E., et al. 2009,\apj, 706, 535
\bibitem[Trouille \& Barger 2010]{trb10} Trouille, L., \& Barger, A. J. 2010, \apj, 722, 212
\bibitem[Ueda et al. 2003]{ued03} Ueda, Y., Akiyama, M., Ohta, K., \& Miyaji, T. 2003, \apj, 598, 886
\bibitem[Vanden Berk et al. 2001]{van01} Vanden Berk, D. E., et al. 2001, \aj, 122, 594
\bibitem[Vaughan et al. 1999]{vau99} Vaughan, S., Reeves, J., Warwick, R., \& Edelson, R. 1999, \mnras, 309, 113
\bibitem[Veilleux \& Osterbrock 1987]{vei87}Veilleux, S., \& Osterbrock, D. E. 1987, \apjs, 63, 295
\bibitem[Vignali et al. 2005]{vig05} Vignali, C., Brandt, W. N., Schneider, D. P., \& Kaspi, S. 2005, \aj, 129, 2519 
\bibitem[Wada et al. 2009]{wad09} Wada, K., Papadopoulos, P. P., \& Spaans, M. 2009, \apj, 702, 63
\bibitem[Wang \& Wei 2006]{wan06} Wang, J., \& Wei, J. Y. 2006, \apj, 648, 158
\bibitem[Wang \& Wei 2008]{waw08}Wang, J., \& Wei, J. Y. 2008, \apj, 679, 86
\bibitem[Wang \& Wei 2010]{waw10}Wang, J., \& Wei, J. Y. 2010, \apj, 719, 1157
\bibitem[Wang et al. 2006]{wah06} Wang, J., Wei, J. Y., \& He, X. T. 2006, \apj, 638, 106
\bibitem[Wang et al. 2004]{wan04} Wang, J. M., Watarai, K. Y., \& Mineshige, S. 2004, \apj, 607, 107
\bibitem[Wang et al. 1996]{wan96}Wang, T., Brinkmann, W., \& Bergeron, J. 1996, \aap, 309, 81
\bibitem[Watabe et al. 2008]{wat08} Watabe, Y., Kawakatu, N., \& Imanishi, M. 2008, \apj, 677, 895
\bibitem[Watson et al. 2009]{wat09} Watson, M. G., et al. 2009, \aap, 493, 339
\bibitem[Wild et al. 2010]{wil11} Wild, V., Heckman, T. M., \& Charlot, S.  2010, \mnras, 405, 933
\bibitem[Wild et al. 2007]{wil07} Wild, V., Kauffmann, G., Heckman, T., et al. 2007, \mnras, 381, 543
\bibitem[Williams et al. 2004]{wil04} 	Williams, R. J., Mathur, S., \& Pogge, R. W. 2004, \apj, 610, 737
\bibitem[Woo \& Urry 2002]{wou02} Woo, J. H., \& Urry, C. M. 2002, \apj, 579, 530
\bibitem[Woo et al. 2012]{woo12} Woo, J. H., Kim, J. H., Imanishi, M., \&  Park, D. 2012, \aj, 143,49
\bibitem[Woo et al. 2010]{woo00} Woo, J. H., et al. 2010, \apj, 716, 269
\bibitem[Worthey \& Ottaviani 1997]{woO97} Worthey, G., \& Ottaviani, D. L. 1997, \apjs, 111, 377
\bibitem[Xu et al. 2003]{xu03} Xu, D. W., Komossa, S., Wei, J. Y., Qian, Y., \& Zheng, X. Z. 2003, \apj, 590, 73
\bibitem[Zamfir et al. 2008]{zam08} Zamfir, S., Sulentic, J. W., \& Marziani, P. 2008, \mnras, 387, 865
\bibitem[Zdziarski et al. 1995]{zdz95} Zdziarski, A. A., Johnson, W. N., Done, C., Smith, D., \& McNaron-Brown, K. 1995, \apjl, 438, 63
\bibitem[Zdziarski et al. 2000]{zdz00} Zdziarski, A. A., Poutanen, J., \& Johnson, W. N. 2000, \apj, 542, 703
\bibitem[Zezas et al. 2003]{zez03} Zezas, A., Ward, M. J., \& Murray, S. S. 2003, \apjl, 594, 31 
\bibitem[Zhou et al. 2005]{zho05} Zhou, H. Y., Wang, T. G., Dong, X. B., Wang, J., \& Lu, H. 2005, Mem. Soc. Astron. Italiana, 76, 93
\bibitem[Zhou et al. 2006]{zho06} Zhou, H. Y., Wang, T. G., Yuan, W. M., Lu, H. L., Dong, X. B., Wang, J. X., \& Lu, Y. J. 2006, \apjs, 166, 128
\bibitem[Zhou \& Zhang 2010]{zhz10} Zhou, X. L., \& Zhang, S. N. 2010, \apjl, 713, 11
\bibitem[Zhou \& Zhao 2010]{zho10} Zhou, X. L., \& Zhao, Y. H. 2010, \apjl, 720, 206
\bibitem[Zycki et al. 1994]{zyc94} Zycki, P. T., Krolik, J. H., Zdziarski, A A., \& Kallman, T. R. 1994, \apj, 437, 597

\end{thebibliography}
\end{document}